\pdfoutput=1

\documentclass[a4paper,12pt]{article}
\usepackage{epsfig,amssymb,amsmath,textcomp,caption,verbatim,gensymb,wrapfig,mathtools}
\usepackage{epstopdf}
\usepackage{graphicx}
\usepackage{amsfonts,braket}
\usepackage[top=1.0in, left=1.0in, bottom=1.5in, right=1.0in]{geometry}
\usepackage[font={small}]{caption}
\usepackage{subfig}
\usepackage[colorlinks=true,citecolor=blue]{hyperref}
\usepackage{cleveref}

\numberwithin{equation}{section}

\newcommand{\Tr}{\mathop{\mbox{\rm Tr}}}
\newcommand{\p}{\partial}
\newcommand{\non}{\nonumber\\}

\begin{document}
	
\thispagestyle{empty}
\renewcommand{\thefootnote}{\fnsymbol{footnote}}
\vskip 3em
\begin{center}
{{\bf \Large The $B=5$ Skyrmion as a two-cluster system}}\\[15mm]
{\large Sven Bjarke Gudnason\footnote{email: bjarke(at)impcas.ac.cn}
 and Chris Halcrow\footnote{email: christ(at)impcas.ac.cn}} \\[1pt]
\bigskip\bigskip
{\it 
Institute of Modern Physics, \\ 
Chinese Academy of Sciences,\\
Lanzhou 730000, China.} \\[20pt]

\vspace{8mm}	
{\bf Abstract}
\end{center}
The classical $B=5$ Skyrmion can be approximated by a two-cluster
system where a $B=1$ Skyrmion is attached to a core $B=4$ Skyrmion. We
quantize this system, allowing the $B=1$ to freely orbit the core. The
configuration space is 11-dimensional but simplifies significantly
after factoring out the overall spin and isospin degrees of
freedom. We exactly solve the free quantum problem and then
 include an interaction potential between the
Skyrmions numerically. The resulting energy spectrum is compared to the
corresponding nuclei -- the Helium-5/Lithium-5 isodoublet. We find approximate parity doubling not seen in the experimental data. In addition, we fail to obtain the correct ground state spin. The framework laid out for this two-cluster system can readily be
modified for other clusters and in particular for other $B=4n+1$
nuclei, of which the $B=5$ is the simplest example.

\vfill
\newpage
\setcounter{page}{1}
\setcounter{footnote}{0}
\renewcommand{\thefootnote}{\arabic{footnote}}

\section{Introduction}
	
The Skyrme model is a nonlinear theory of pions which admits topologically non-trivial configurations called Skyrmions, labeled by a topological charge $B$. Skyrme's pioneering idea \cite{Sky} was to identify these configurations with nuclei and the topological charge with baryon number.

The theory captures many phenomenological features of nuclei such as isospin symmetry, $\alpha$-clustering \cite{BMS}, rotational bands \cite{MMW} and even contains a version of the liquid drop model \cite{ASW}. Recently the model has successfully explained the energy spectrum of Carbon-12 and the large root-mean-square charge radius of the Hoyle state \cite{LM}, and has been able to describe the rich energy spectrum of Oxygen-16 \cite{HKM}. The Skyrme model is attractive theoretically as there are few adjustable parameters in the Lagrangian. Further, all nucleon interactions and dynamics are determined by this initial Lagrangian.

The model can approximately reproduce the low-energy spectrum of all light ($B=1-8$) nuclei except the $^{5}$Li/$^5$Be isodoublet \cite{MMW,Hal}. These two nuclei are usually described in the shell model as an
inert core (the $\alpha$-particle) plus one orbiting nucleon. In the
most basic shell model the additional nucleon has either spin
$1/2$ or $3/2$. The spin-orbit interaction breaks this
degeneracy, making the spin $3/2$ state energetically
favored. Hence the ground states of Helium-5 and Lithium-5 have
spin $3/2$. This story is partially mirrored in the Skyrme model. Here, the $B=5$ Skyrmion is approximately described by a $B=4$ Skyrmion plus an additional $B=1$ Skyrmion. In the standard Skyrme model the clusters merge into a $D_2$-symmetric Skyrmion, although it takes little energy to separate them. In modified Skyrme models such as the loosely bound Skyrme model \cite{Gud,Gud2,Gud3} and lightly bound Skyrme model \cite{GHS,GHKMS}, the $B=5$ Skyrmion is very well approximated by the two-cluster system. There should then be a low energy manifold of configurations: those where the $B=1$ orbits the $B=4$ core. Taking account of these degrees of freedom allows us to describe the $B=5$ as a $4+1$ cluster system, just like the shell model. The shell model notion of the spin-orbit force is not present in the Skyrme model; instead there is an interaction potential which depends on the internal orientations of the Skyrmion clusters. The link between these ideas was explored in a $2$-dimensional toy model in \cite{HM}.

In this paper, we consider the quantization of the $B=5$ Skyrmion as a two-cluster system. Each cluster is individually allowed to rotate and isorotate. This is the first time such a system has been quantized within the Skyrme model. The resultant spectrum contains the low energy spin $3/2^-$ and $1/2^-$ states, though in the wrong order. We also find parity doubling, not seen in experimental data. The unwanted $3/2^+$ and $1/2^+$ states are not allowed by the $D_2$-symmetric Skyrmion, which is not included in our configuration space. Its existence should still affect the energy of the states, though the size of this impact is difficult to measure. 

Many nuclei can be described as a core + particle system, such as those close to magic nuclei. The $B=5$ is the simplest of these systems. This motivates us to study the system carefully and the framework we develop should apply more broadly, with a few simple modifications. In fact, much of our formalism applies to any two-cluster system. These are used frequently in nuclear physics, from modeling bound nuclei such as Lithium-7 \cite{TWP} to describing scattering between $\alpha$-particles \cite{ELR} or Carbon-12 nuclei \cite{DW}. Hence we believe our work provides an important step towards understanding this wide range of problems within the Skyrme model.

In the next Section we carefully set up our model of the $^5$He/$^5$Li isodoublet as a $4+1$ two-cluster system within the Skyrme model. We then quantize the system in Section \ref{sec:3} before dealing with the cubic symmetry of the core in Section \ref{sec:4}. The methods described in this section are very general and can be applied to any vibrational quantization in the Skyrme model. We then find states with definite parity and discuss some additional symmetries of the configuration space in Section \ref{sec:5}. The energies of the derived wavefunctions are calculated and compared to experimental data in Section \ref{sec:6}. The framework we develop in this paper may be applied to a wide range of systems: we suggest several avenues for further work in Section \ref{sec:7}. Concluding remarks can be found in Section \ref{sec:8}.
	
\section{The Skyrme model, our configuration space and relative coordinates} \label{sec:2}

The variant of the Skyrme model we consider in this paper is the
most commonly used version, albeit with a pion mass and the
possibility of a loosely bound potential \cite{Gud}. The Lagrangian density is
\begin{align} \label{Lagrangian}
\mathcal{L}
= \Tr\left[
\frac{1}{2}L_\mu L^\mu
+\frac{1}{16}[L_\mu,L_\nu][L^\mu,L^\nu]
-m_1^2(\mathbf{1}_2-U)
\right]
-\frac{1}{4}m_2^2\left(\Tr[\mathbf{1}_2-U]\right)^2,
\end{align} 
where $L_\mu\equiv U^\dag\p_\mu U$ is the left-invariant
$\mathfrak{su}(2)$-valued current, $U$ is the Skyrme field, $m_1$ is
the pion mass and $m_2$ is a parameter of the loosely bound
potential. The Skyrme field is $SU(2)$-valued and can be written in terms of the
pion field $\boldsymbol{\pi}(\boldsymbol{x})$ as
\begin{equation}
U = \mathbf{1}_2\sigma + i\boldsymbol{\tau}\cdot\boldsymbol{\pi}\,,
\end{equation}
where $\boldsymbol{\tau}$ is the triplet of Pauli matrices and $\sigma$ is an auxiliary field which ensures that $U$ takes values in $SU(2)$. To visualize a Skyrme configuration we plot a contour of constant
energy density. This is then colored to represent the direction of
the pion field at that point on the energy contour. 
The Skyrme field is colored white/black when 
$\hat{\pi}_3 = \pm 1$ and red, green and blue when
$\hat{\pi}_1+i\hat{\pi}_2 = \exp(0)$, $\exp(2i \pi/3)$ and
$\exp(4 i \pi /3)$ respectively, where
$\hat{\boldsymbol{\pi}}\equiv\boldsymbol{\pi}/|\boldsymbol{\pi}|$ is
the normalized pion field.

\begin{figure}[!ht]
\begin{center}
\mbox{\subfloat[$m_2=0$]{\includegraphics[width=0.2\linewidth]{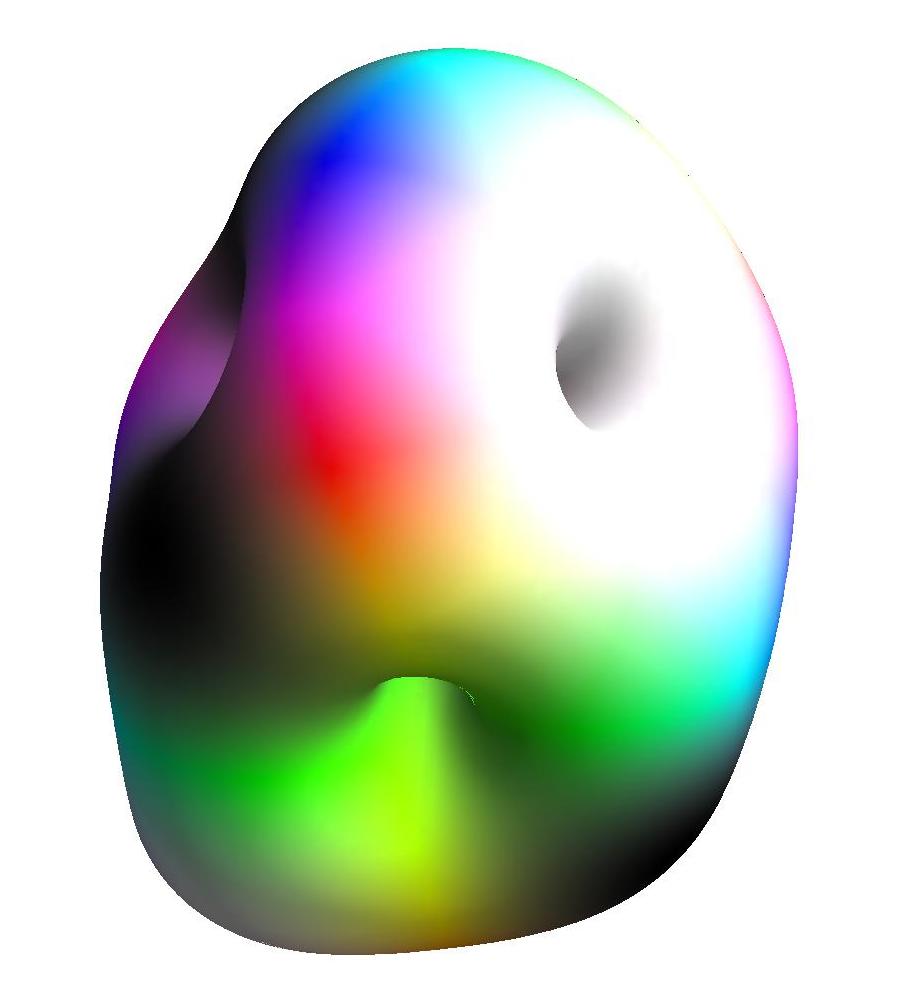}}
\subfloat[$m_2=0.8$]{\includegraphics[width=0.2\linewidth]{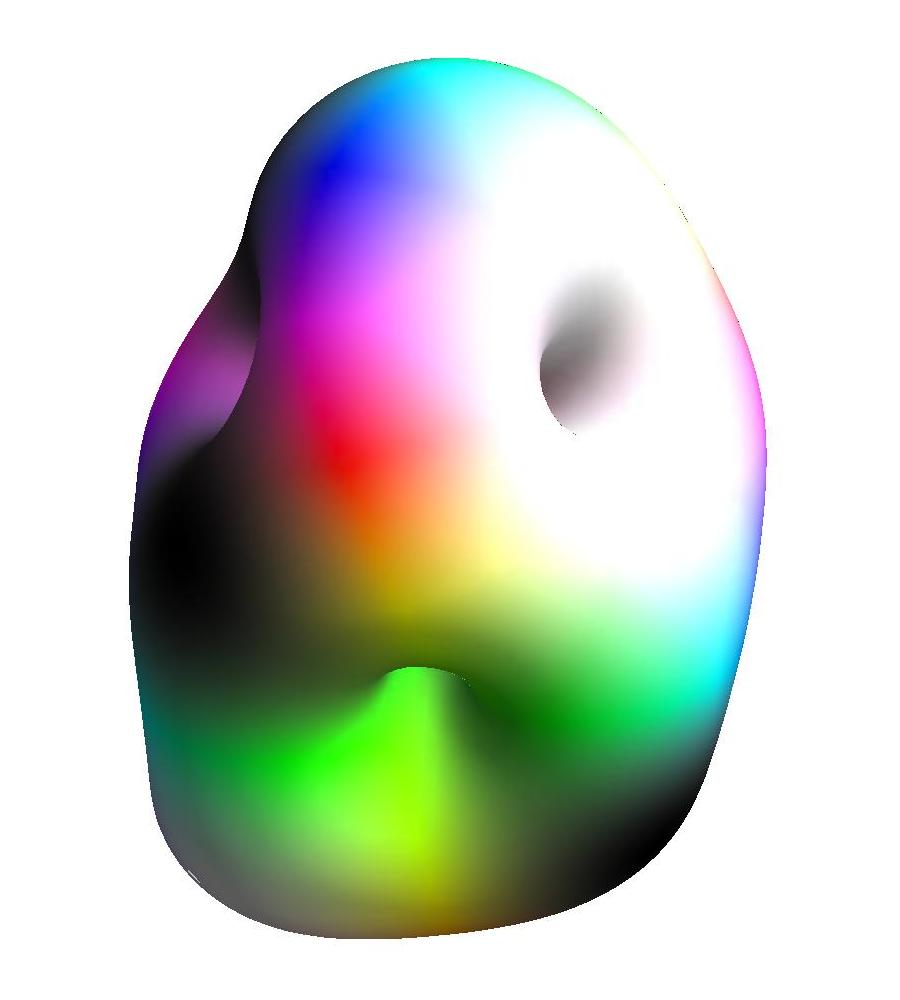}}
\subfloat[$m_2=1.6$]{\includegraphics[width=0.2\linewidth]{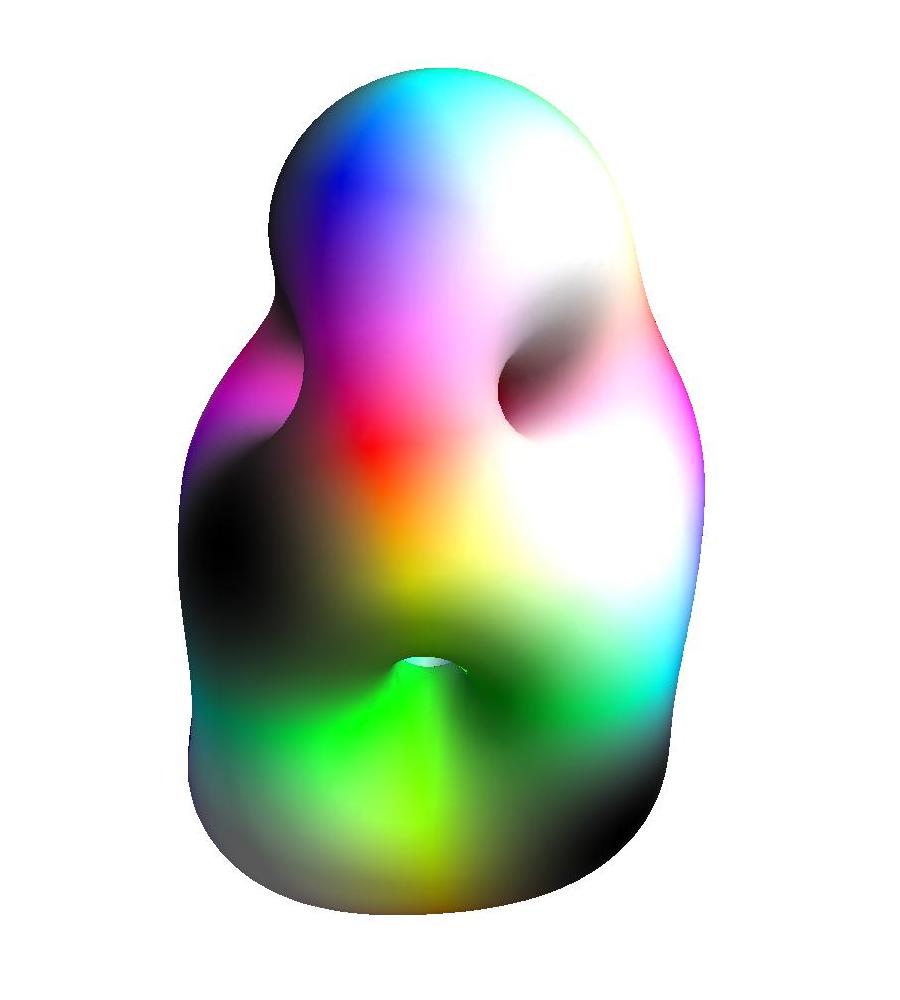}}
\subfloat[$m_2=2.4$]{\includegraphics[width=0.2\linewidth]{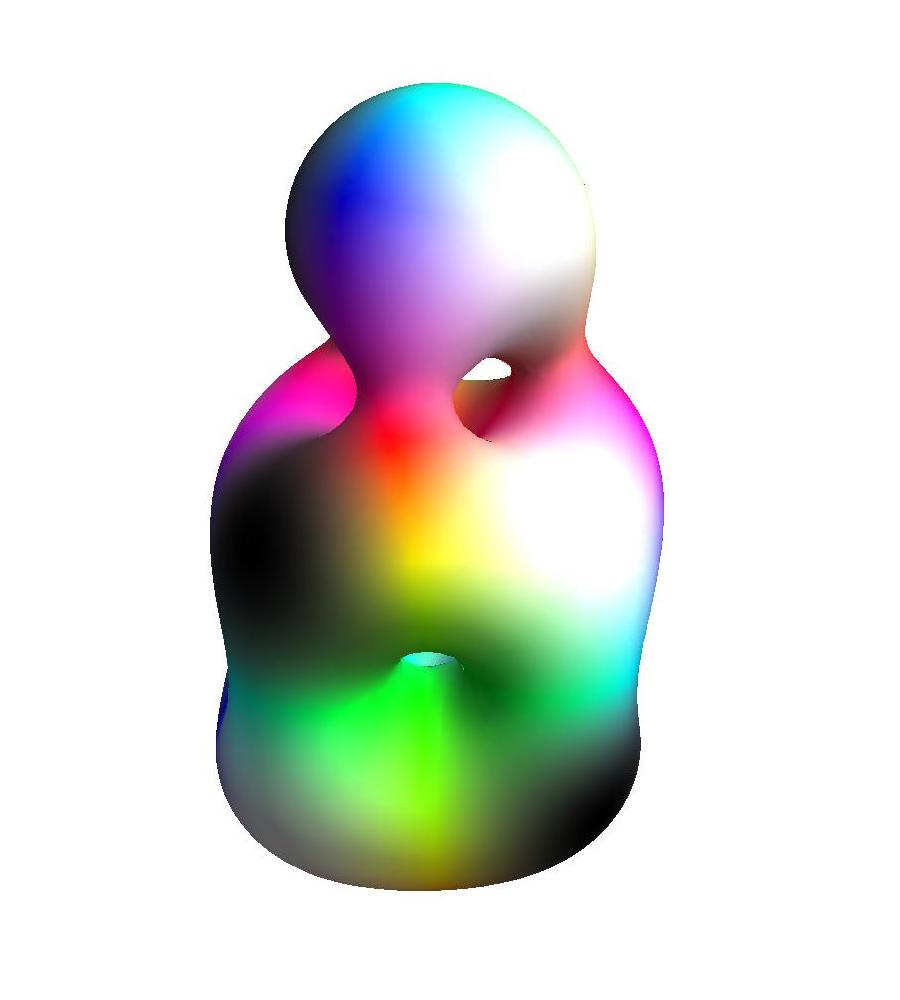}}
\subfloat[$m_2=3.2$]{\includegraphics[width=0.2\linewidth]{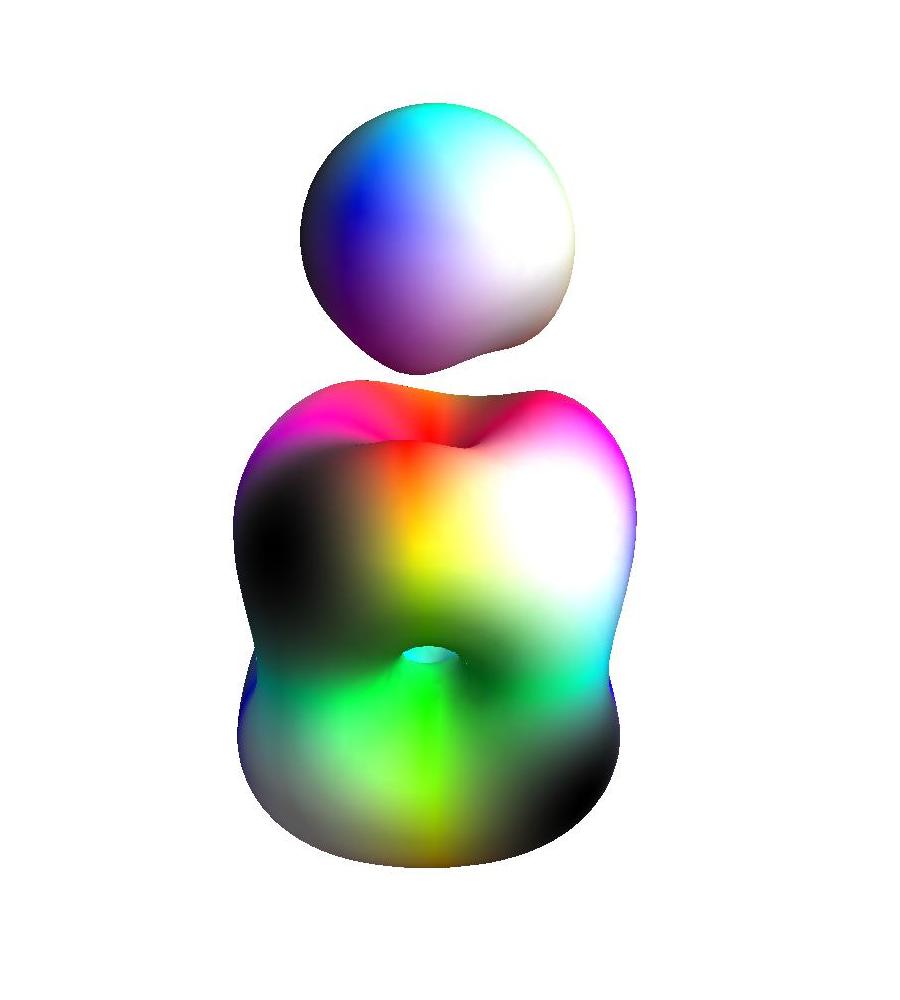}}}
\caption{The $B=5$ Skyrmion with increasing value of the loosely bound
potential parameter $m_2$. The leftmost figure corresponds to the standard Skyrme
model, while the rightmost figure is the loosely bound Skyrme model
with the $B=1$ Skyrmion detached from the cube. The pion mass is taken to
be $m_1=1$. }
\label{fig:m2pot}
\end{center}
\end{figure}

A Skyrmion is a static energy minimizer of \eqref{Lagrangian}, for a given topological charge $B$. When the loosely bound potential \cite{Gud,Gud2} is turned off, $m_2=0$, the $B=5$ Skyrmion has $D_{2}$ symmetry. As $m_2$ is increased, the Skyrmion separates into two clusters. A $B=1$ Skyrmion is gradually detached from a $B=4$ core which acquires approximate cubic symmetry. This process is shown in Figure \ref{fig:m2pot}.

\pagebreak

\begin{wrapfigure}{r}{0.5\textwidth}
\centering
\includegraphics[width=0.3\textwidth]{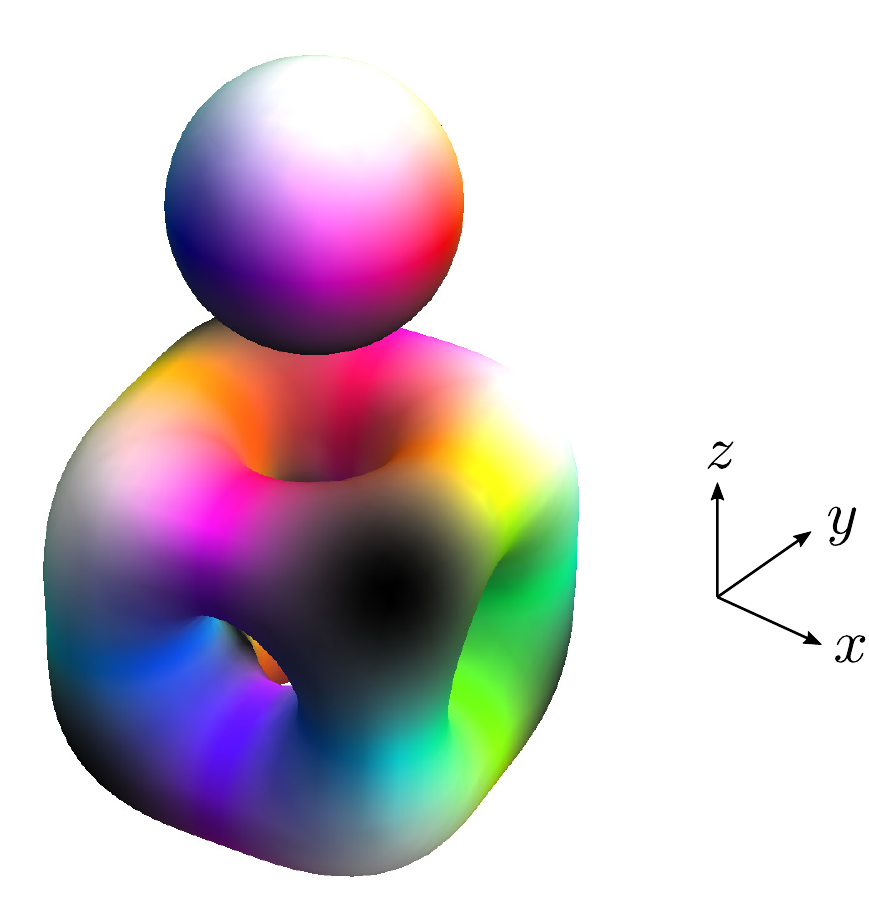} 
\caption{The basic set up for our model. A $B=1$ Skyrmion orbits a
  $B=4$ Skyrmion. }
\label{SetUp}
\end{wrapfigure} 
	
The initial set-up we study in this paper is based on the two-cluster
system displayed in Figure \ref{SetUp}. Here, a $B=1$ Skyrmion orbits
a cubic $B=4$ Skyrmion at a fixed radius. Both clusters are free to
rotate, isorotate and move around one another. We take the $B=4$ to be
fixed at the origin. Although we are not in the center of mass frame,
several other aspects of the problem simplify using this
approximation.  There are different ways to interpret the model. One is
that the configurations, such as the one in Figure \ref{SetUp}, are
good approximations to the classical low energy configurations of the
model. This is true for the loosely bound model, as we have seen in Figure \ref{fig:m2pot}. Another way to
interpret the model is that we are simply using convenient coordinates
to describe a submanifold of the $B=5$ vibrational manifold. In this
case, the image in Figure \ref{SetUp} does not accurately describe the
configuration represented by this point in configuration
space. Instead, the point corresponds to some deformed version of the configuration, where the clusters have merged and distorted. Regardless of interpretation, our
analysis of the symmetries and quantization of the system will
apply. However, the details of the specific moments of inertia and
energies will change.  

Our plan is to solve the free system where the two clusters do not
interact (though are bound together since we have fixed their
separation). We will then include the interaction energy by
diagonalizing a truncated set of free states with respect to the
Hamiltonian, including a numerically generated potential term. This
method breaks down if the potential is strong and hence we are solving
the problem in the weak-coupling limit. We will suggest how one may
study the strong coupling limit in Section \ref{sec:7}. The system is described
by eleven coordinates: there are six which specify the orientation of the
$B=4$ cluster in space and isospace, three describe the $B=1$'s
orientation and the remaining two specify the orbital position of
the $B=1$. This is a formidable number of degrees of freedom. However,
the system transforms simply under overall rotations and
isorotations. In particular, the interaction potential is invariant
under such transformations. If we can ``factor out" these symmetries
only five coordinates, describing the relative interaction between
clusters, will remain. This type of reduction is common in the study
of comets, many of which are described by two-cluster systems bound
together by the gravitational force. We closely follow the work of
Maciejewski \cite{Mac} who first solved such a problem. 

To proceed we must introduce notation for the individual momenta of the Skyrmions. We define
\begin{itemize}
	 \setlength\itemsep{0.1em}
	\item $\boldsymbol{J}^{(i)}$: space-fixed angular momentum of the $B=i$ Skyrmion.
	\item $\boldsymbol{L}^{(i)}$: body-fixed angular momentum of the $B=i$ Skyrmion.
	\item $\boldsymbol{I}^{(i)}$: space-fixed isospin of the $B=i$ Skyrmion.
	\item $\boldsymbol{K}^{(i)}$: body-fixed isospin of the $B=i$ Skyrmion.
	\item $\boldsymbol{R}$: position of the $B=1$ Skyrmion relative to the $B=4$, in the space-fixed frame.
	\item $\boldsymbol{P}$: orbital angular momentum of the $B=1$ Skyrmion relative to the $B=4$ Skyrmion.
\end{itemize}
The body-fixed and space-fixed angular and isoangular momenta are related by the orthogonal matrices $A^{(i)}$ and $B^{(i)}$, also known as the attitude matrices. Explicitly
\begin{equation}
\boldsymbol{J}^{(i)} = A^{(i)} \boldsymbol{L}^{(i)} \, \text{ and } \,  \boldsymbol{I}^{(i)} = B^{(i)} \boldsymbol{K}^{(i)} \, .
\end{equation} 
Further, the momenta of the $B=1$ are related. We use the conventions that the body-fixed momenta are related as $ \boldsymbol{L}^{(1)}=-\boldsymbol{K}^{(1)}$. This is the convention used in \cite{MMW} but is opposite to that used in \cite{Kru}. It implies that the space-fixed momenta obey
\begin{equation}
\boldsymbol{J}^{(1)} = -A^{(1)}B^{(1)T}\boldsymbol{I}^{(1)} \, .
\end{equation}
One may combine $A^{(1)}B^{(1)T}$ into a single matrix, reflecting the fact that the $B=1$'s orientation can be parametrized by only three coordinates . We do not do this, to keep the symmetries of the formalism explicit and to emphasize that this method may be applied to any two-cluster system, not just a core + particle system.

The total (iso)angular momentum is the sum of the individual (iso)momenta and the orbital momentum. A simple way to describe these quantities is to go into the body-fixed frame of the $B=4$ Skyrmion. In this frame, the total angular momentum is
\begin{equation}
\boldsymbol{J} = \boldsymbol{L}^{(4)} + A^{(4) T}A^{(1)}\boldsymbol{L}^{(1)}+A^{(4) T}\boldsymbol{l} \, , \label{Jdef}
\end{equation} 
where $\boldsymbol{l} = \boldsymbol{R}\times \boldsymbol{P}$ is the orbital angular momentum in the space-fixed frame. The total isoangular momentum (isospin) is
\begin{align} \label{Idef}
\boldsymbol{I} &= \boldsymbol{K}^{(4)} + B^{(4) T}B^{(1)}\boldsymbol{K}^{(1)} \, .
\end{align}
 As these are body-fixed momenta, they will satisfy the anomalous commutation relations when quantized. These are
\begin{equation}
[J_i, J_j] = -i\epsilon_{ijk}J_k \, , \, \, \, \, [I_i, I_j] = -i\epsilon_{ijk}I_k \label{JIcomms}\,   \text{ and } \, [J_i, I_j]=0  \, .
\end{equation}
After quantization, the conserved quantities $J$ and $I$ will correspond to the spin and isospin of the nucleus.

We can now define the relative momenta. One has some choice in how to define these and ours are chosen for their simplicity in describing the Finkelstein-Rubinstein constraints which we meet in the next Section. The relative momenta are
\begin{align} \label{STdef}
\boldsymbol{S} &= A^{(4) T}A^{(1)}\boldsymbol{L}^{(1)}+A^{(4) T}\boldsymbol{l} \, ,\\
\boldsymbol{T} &=  B^{(4) T}B^{(1)}\boldsymbol{K}^{(1)} \, . 
\end{align}
These act on the $B=1$ Skyrmion, leaving the cubic core unchanged. The momentum $\boldsymbol{S}$ generates a rolling motion of the $B=1$ around the $B=4$ while $\boldsymbol{T}$ acts only on the orientation of the $B=1$. They satisfy the usual commutation relations
\begin{equation}
[S_i, S_j] = i\epsilon_{ijk}S_k \, , \, \, \, \, [T_i, T_j] = i\epsilon_{ijk}T_k \, , \label{STcomms}
\end{equation}
and commute with each other and the overall momenta $\boldsymbol{J}$ and $\boldsymbol{I}$.  One could think of the relative coordinates as vibrational degrees of freedom of the $B=5$ Skyrmion. Then, in the language of \cite{HKM}, $\boldsymbol{S}$ and $\boldsymbol{T}$ generate the vibrational manifold while $\boldsymbol{J}$ and $\boldsymbol{I}$ generate the rotational manifold. 

 The classical configuration space of the system is naively 
\begin{equation}
\mathcal{M} = SO(3)_J \times SO(3)_I \times \frac{ SO(3)_S \times SO(3)_T}{U(1)} \, .
\end{equation}
The $U(1)$ factor accounts for the fact that there is a degeneracy in the $S/T$ space. This is clear as we only require five degrees of freedom to describe the position and orientation of the $B=1$ but $SO(3)_S\times SO(3)_T$ is six-dimensional. Explicitly, a simultaneous rotation about the $3$-axis in the $S$ and $T$ spaces leaves the system invariant. We have also not yet accounted for the discrete symmetries of the system. We consider both the $U(1)$ symmetry and the discrete symmetries carefully in the next Section.

\section{Quantization} \label{sec:3}

Having described the classical configuration space of our model, we can now quantize the system using a semi-classical quantization. This is done by promoting the coordinates on $\mathcal{M}$ to dynamical degrees of freedom. Each classical momentum becomes a quantum operator and the four momenta produce quantized spins which we denote $J, I, S$ and $T$. A general wavefunction can be written as
\begin{equation}
\Ket{\Psi} = \sum_{j_3, i_3, s_3, t_3, s_3', t_3'} a_{j_3 i_3 s_3 t_3 s_3' t_3'} \Ket{J\, j_3\, J}\Ket{I \, i_3\, I}\Ket{S\, s_3 \, s_3'}\Ket{T\, t_3\, t_3'} \, . \label{wavefn}
\end{equation}
where the three labels of each spin state represent the total spin, the projection onto the body-fixed third axis and the projection onto the space-fixed third axis. For the overall angular momentum and isospin ($J$ and $I$), the space-fixed projections do not affect the structure of the spin state or its energy. Hence we may set them to any value and we choose to fix them equal to their total spins. However, the space-fixed projections for the relative momenta do alter the energies and so we must allow for linear combinations of these, as well as the body-fixed projections.

The $U(1)$ degeneracy in the $S/T$ part of the configuration space
$\mathcal{M}$ constrains the wavefunction \eqref{wavefn}. To see how,
we put coordinates on this part of the space. These describe the
relative orientation between the $B=1$ and $B=4$ Skyrmions. The
rolling motion (generated by $S$) may be parametrized using Euler
angles $(\phi, \theta, \psi)$. We will use the passive Z-Y-Z
conventions (applying $\psi$ first, then $\theta$ and finally
$\phi$). Using these, when $\psi=0$, the remaining coordinates are the
usual spherical polars. The internal motion of the $B=1$ (described by
$T$) can also be parametrized by Euler angles $(\alpha, \beta,
\gamma')$. The spin states can be written in terms of the Euler angles
using Wigner-D functions. The relations are 
\begin{equation}
\Ket{T\, t_3\, t_3'} = D^T_{t_3' t_3}(\alpha, \beta, \gamma') \quad
\text{and} \quad \Ket{S\, s_3 \, s_3'} = D^S_{s_3' s_3}(\phi, \theta, \psi) \, ,\label{Swv} 
\end{equation}
where
\begin{equation}
D^T_{t_3' t_3}(\alpha, \beta, \gamma') = e^{ i \alpha t_3} D^T_{t_3' t_3}(0,\beta,0)e^{i \gamma' t_3'} \, .
\end{equation}
The system is invariant under a simultaneous \emph{increase} in both
$\gamma'$ and $\psi$, as these coordinates rotate the $B=1$
Skyrmion around the $z$-axis in opposite directions.
This leads to a constraint on the space-fixed projections of $S$ and $T$
\begin{equation}
\left(\partial_{\gamma'} + \partial_\psi \right)\Ket{\Psi} = \Ket{\Psi} \implies  s_3' = {-t_3'} \, ,
\end{equation}
and so the overall wavefunction contains a factor
\begin{equation}
e^{i s_3'\psi +i t_3'\gamma'} = e^{i t_3' (-\psi+\gamma')} \equiv e^{i t_3' \gamma} \, ,
\end{equation}
where we have defined a new coordinate $\gamma$. Using this coordinate is equivalent to setting $\psi=0$ and $\gamma'=\gamma$ in \eqref{Swv}. Hence the general wavefunction is given by
\begin{equation}
\Ket{\Psi} = \sum_{j_3, i_3, s_3, t_3, t_3'} a_{j_3 i_3 s_3 t_3 t_3'} \Ket{J\, j_3\, J}\Ket{I \, i_3\, I}D^S_{{-t_3'} s_3}(\phi, \theta,0) D^T_{t_3' t_3}(\alpha, \beta, \gamma) \, . \label{wavefn2}
\end{equation}
Note that $|t_3'|$ can only take values up to $\text{min}(2 S+1, 2T+1)$.

\begin{figure}[!ht]
	\centering
	\includegraphics[width=0.9\textwidth]{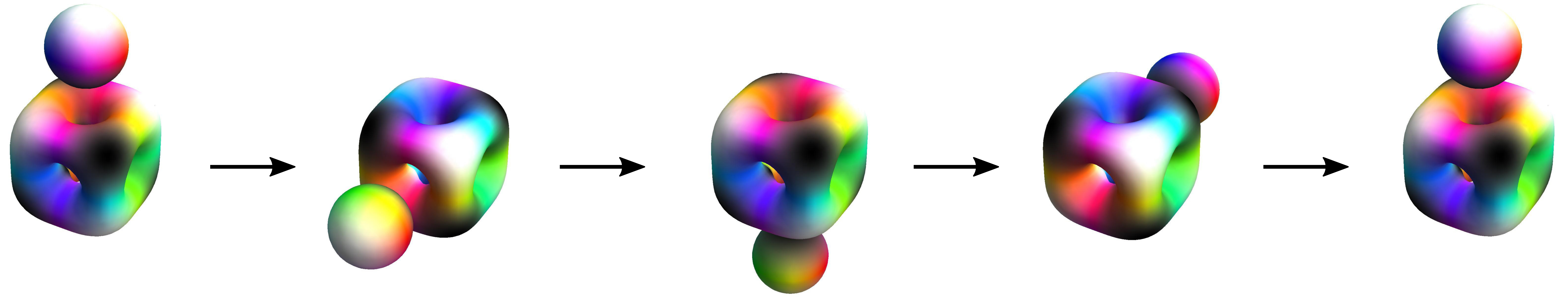} 
	\caption{A $2\pi$ rotation around the $x$-axis. Note that the $B=1$ Skyrmion does a $2\pi$ rotation on its own axis and so the wavefunction should pick up a minus sign under this rotation. } \label{Rotation}
\end{figure}

The Finkelstein-Rubinstein (FR) constraints \cite{FR} deal with the fact that the $B=5$ Skyrmion is constructed out of five nucleons, and so the system must obey fermionic statistics. The constraints provide restrictions on the allowed wavefunctions and determine if $J, I, S$ and $T$ are integers or half-integers. This may be determined by considering the rotations physically. Consider a $2\pi$ rotation about the 1-axis in the $J$ space, as seen in Figure \ref{Rotation}. This rotates the $B=4$ Skyrmion by $2\pi$ and the $B=1$ Skyrmion rolls around the cube in a circular orbit. Overall, both Skyrmions rotate by $2\pi$ on their own 1-axes. Physically the $B=1$ represents a single nucleon, a fermion, while the $B=4$ represents four nucleons, a boson. Hence the wavefunction should transform by $-1$ under this rotation. Hence $J$ must be a half-integer. This was already guaranteed by the fact that $B$ is odd. Similar arguments show that the other conserved spins ($I$, $S$ and $T$) must each be half-integers. Note that all of the spins are conserved in the free system but after inclusion of the potential only $J$ and $I$ will be. States with different $S$ and $T$ values will mix in the non-free theory. The classical configuration space $\mathcal{M}$ cannot be used to describe systems with half-integer spins. Finkelstein and Rubinstein showed that we must use the double cover of $\mathcal{M}$ instead, and so the quantum configuration space is
\begin{equation} 
\mathcal{M}_q = SU(2)_J \times SU(2)_I \times \frac{SU(2)_S \times SU(2)_T}{U(1)} \, .
\end{equation}

FR constraints also apply to the discrete symmetries of the configuration space. These are inherited from the discrete symmetries of the clusters. The $B=1$ has no discrete symmetries (and we have already dealt with its continuous symmetry by our parametrization) but the $B=4$ has cubic symmetry. The wavefunction either transforms trivially or picks up a sign under each element of the symmetry group. To determine which, we may use the algorithm developed in \cite{Kru}. The explicit symmetries are most easily expressed in terms of the cube's body-fixed momenta. In these terms, the group is generated by a $C_4$ symmetry
\begin{equation}
\text{exp}\left(- \frac{\pi i}{2} \hat{L}^{(4)}_3  -  \pi i \hat{K}^{(4)}_1\right)\Ket{\Psi} = \Ket{\Psi}
\end{equation}
and a $C_3$ symmetry
\begin{equation}
\text{exp}\left(- \frac{2\pi i}{3 \sqrt{3}} \left(\hat{L}^{(4)}_1+\hat{L}^{(4)}_2+\hat{L}^{(4)}_3\right)  -  \frac{2\pi i}{3} \hat{K}^{(4)}_3\right)\Ket{\Psi} = \Ket{\Psi} \, .
\end{equation}
To find out how these operators act in $\mathcal{M}$ they must be expressed in terms of the new momenta. Our choice of $\boldsymbol{S}$ and $\boldsymbol{T}$ means that the relation is simple. The constraints become
\begin{align}
\text{exp}\left(- \frac{\pi i}{2} (\hat{J}_3-\hat{S}_3)  -  \pi i (\hat{I}_1-\hat{T}_1)\right)\Ket{\Psi} &= \Ket{\Psi} \, ,  \label{FRs2} \\
\text{exp}\left(- \frac{2\pi i}{3 \sqrt{3}} \left(\hat{J}_1+\hat{J}_2+\hat{J}_3-\hat{S}_1-\hat{S}_2-\hat{S}_3\right)  -  \frac{2\pi i}{3} (\hat{I}_3-\hat{T}_3)\right)\Ket{\Psi} &= \Ket{\Psi} \, . \label{FRs1}
\end{align}
The form of these operators highlight the fact that applying a
rotation to the $B=4$ is equivalent to applying a rotation to the
entire two-cluster system, then ``undoing" this rotation for the
$B=1$. The wavefunctions of the form \eqref{wavefn2} which satisfy
\eqref{FRs2} and \eqref{FRs1} are allowed by the symmetries of our
system. We call these permissible wavefunctions. For a fixed $J$, $I$,
$S$ and $T$ these constraints become a simple linear algebra problem
for the coefficients $a$ of \eqref{wavefn2}, which one may solve. We
will take an alternative approach by using representation theory. This
will highlight the similarity between our calculation and other
calculations in many subjects such as the Skyrme model \cite{HKM},
molecular vibrations \cite{WDC} and other nuclear models
\cite{Den}. Additionally, the method laid out here  may be applied to any
rotational-vibrational system in the Skyrme model. A reader less
interested in our rather technical calculation may wish to skip the
next Section. 

Before proceeding we comment on the exponentiation of the rotation operators, used in the FR constraints. To evaluate these operators we must find a matrix representation of the $\mathfrak{su}(2)$ Lie algebras. The $S$ and $T$ operators obey the usual commutation relations \eqref{STcomms} and so the matrix representation $\boldsymbol{M}$ for the operators $\hat{\boldsymbol{S}}$ and $\hat{\boldsymbol{T}}$ are well known. For example, the spin $1/2$ matrices are simply the Pauli matrices (divided by one-half) and satisfy
\begin{equation}
[M_i,M_j] = i \epsilon_{ijk}M_k \, .
\end{equation}
However, the $J$ and $I$ operators satisfy the \emph{anomalous} commutation relations \eqref{JIcomms}. Hence, we cannot use the usual $\boldsymbol{M}$. Instead we use the conjugate representation $\boldsymbol{M}^*$ which satisfies
\begin{equation}
[M^*_i,M^*_j] = -i \epsilon_{ijk}M^*_k \, .
\end{equation}
For example, a $\tfrac{\pi}{2}$ rotation about the $(0,1,0)$ axis acts on the $S=1/2$ wavefunctions as
\begin{equation}
\exp\left(-\tfrac{ \pi i}{2}M_2\right) = \exp\left(-\frac{  \pi
  i}{4}\begin{pmatrix}0 & i \\ -i & 0 \end{pmatrix} \right) =
\frac{1}{\sqrt{2}}\begin{pmatrix}1 & 1 \\ -1 & 1 \end{pmatrix} \, ,
\end{equation}
but the same rotation acts on a $J=1/2$ wavefunction as
\begin{equation}
\exp\left(-\tfrac{ \pi i}{2}M^*_2\right) = \exp\left(-\frac{ \pi i}{4}\begin{pmatrix}0 & -i \\ i & 0 \end{pmatrix} \right) = \frac{1}{\sqrt{2}}\begin{pmatrix}1 & -1 \\ 1 & 1 \end{pmatrix} \, .
\end{equation}

\section{Constructing permissible wavefunctions} \label{sec:4}

In this Section we construct a basis of wavefunctions which satisfy the constraints \eqref{FRs2} and \eqref{FRs1} by splitting the total symmetry group into two parts. One which acts only on the $J$/$I$ space; the other on the $S$/$T$ space. Suppose we have a basis of states in the $J/I$ space $\{ \ket{\Psi_i} \}$ and a similar basis in the $S/T$ space $\{ \Phi_i \}$. We may act on the $\ket{\Psi_i}$ states using the $J$/$I$ part of one of the operators which feature in \eqref{FRs2} and \eqref{FRs1}. For example, the $C_4$ operator transforms the state as
\begin{equation}
\Ket{\Psi_i} \to e^{-\tfrac{\pi i}{2}\hat{J}_3 - \pi i \hat{I}_1 }\Ket{\Psi_i} = N_{ij}\Ket{\Psi_j} \, ,
\end{equation}
for some matrix $N$. A total wavefunction
\begin{equation}
\Ket{\Psi} = \sum_i \Phi_i \ket{\Psi_i}
\end{equation}
will be invariant under \eqref{FRs2} and \eqref{FRs1} if the $ \Phi_i $ transform as
\begin{equation}
\Phi_i \to e^{\tfrac{\pi i}{2}\hat{S}_3 + \pi i \hat{T}_1 }\Phi_i \to N_{ji}^{-1}\Phi_j = N_{ij}^* \Phi_j\, . \label{phitrans}
\end{equation}
That is, if the $ \Phi_i $ transform as the dual representation compared to the $\Ket{\Psi_i}$\footnote{We thank Jonathan Rawlinson for bringing this to our attention.}. For the unitary groups, the dual matrix representations are simply complex conjugates of each other. 

To construct permissible wavefunctions, we must first understand the representations of the symmetry groups, which we denote $\mathcal{G}_{JI}$ and $\mathcal{G}_{ST}$. First we consider $\mathcal{G}_{JI}$, which is generated by the $J/I$ part of the full symmetry transformations \eqref{FRs2} and \eqref{FRs1}. These are
\begin{align}
&\hat{C}_4^{J,I} =\text{exp}\left(-\frac{\pi i}{2} \hat{J}_3 - \pi i \hat{I}_1\right) \, ,  \nonumber \\
&\hat{C}_3^{J,I} = \text{exp}\left(- \frac{2\pi i}{3 \sqrt{3}} \left(\hat{J}_1+\hat{J}_2+\hat{J}_3\right)  -  \frac{2\pi i}{3} \hat{I}_3\right)\, .  \label{JIsymm}
\end{align}
The group is closely related to the cubic group $O$. The difference can be seen by considering the $C_4$ element. Applying this four times gives a $2\pi$ rotation and a $4\pi$ isorotation. Hence, the wavefunction should pick up an overall minus sign under $(C_4)^4$ meaning that $(C_4)^4 \neq e $, where $e$ is the identity operator. Thus $\mathcal{G}_{JI}$ is not homomorphic to $O$. In contrast, $(C_3)^3 = e$. Overall, the group has $48$ elements though we neglect the inversion operator for now. The character table of $\mathcal{G}_{JI}$ is displayed in Table \ref{CharTable}. It contains the character table of $O$ since the quotient of of $\mathcal{G}_{JI}$ by $2\pi$ isorotations is $O$. There are three irreducible representations (irreps) not contained in $O$. A four-dimensional irrep, denoted $\boldsymbol{4}$, descends from the fundamental representation of $SU(2)_J \times SU(2)_I$. The remaining irreps both have dimension two and so we label them $\boldsymbol{2^+}$ and $\boldsymbol{2^-}$. The symmetry group $\mathcal{G}_{ST}$ is isomorphic to $\mathcal{G}_{JI}$ and hence they share the same character table.

\bgroup
\def \arraystretch{1.2}
\begin{table}[!ht]
	\begin{center}
		\begin{tabular}{| c | c | c | c | c | } \hline
			Irrep. & $e$ & $\hat{C}_4$ & $\hat{C}_3$ & $2\pi$ rotation \\ \hline
			$\boldsymbol{1^+}$ & 1 & 1 & 1 & 1 \\ 
			$\boldsymbol{1^-}$ & 1 & ${-1}$ & 1 & 1 \\
			$\boldsymbol{2^{\,0}}$ & 2 & 0 & ${-1}$ & 2 \\
			$\boldsymbol{3^+}$ & 3 & 1 & 0 & 3 \\
			$\boldsymbol{3^-}$ & 3 & ${-1}$ & 0 & 3 \\ \hline
			$\boldsymbol{2^+}$ & 2 & $\sqrt{2}i$ & ${-1}$ & ${-2}$ \\
			$\boldsymbol{2^-}$ & 2 & ${-\sqrt{2}}i$ & ${-1}$ & ${-2}$ \\
			$\boldsymbol{4}$ & 4 & 0 & 1 & ${-4}$ \\ \hline
			
		\end{tabular}
		\vskip 7pt
		\caption{The character table for $\mathcal{G}_{JI}$ and $\mathcal{G}_{ST}$. The first five rows show the character table for the usual cubic group.}
		\label{CharTable}
	\end{center}
\end{table}
\egroup

The character table may be used to decompose the set of spin states
\begin{equation}
\{ \Ket{J\, j_3\, J}\Ket{I \, i_3\, I} |\,\,  j_3 = -J, \ldots, J\, , \, i_3 = -I, \ldots, I \}
\end{equation}
 into irreducible parts. As an example, take the $J=3/2$, $I=1/2$ basis. First, we find the matrix representation of the operators \eqref{JIsymm} using $\boldsymbol{M}^*$, as explained in Section \ref{sec:3}. This is an 8-dimensional representation which satisfies 
 \begin{equation}
 \text{Tr}\left(C_3^{J=\frac{3}{2},\, I=\frac{1}{2}}\right) = -1 \quad \text{and} \quad \text{Tr}\left(C^{J=\frac{3}{2}, \, I=\frac{1}{2}}_4\right) = 0 \, .
 \end{equation}
  Comparing these traces to the character table, one finds that this representation contains a single copy of $\boldsymbol{4}$, $\boldsymbol{2^+}$ and $\boldsymbol{2^-}$. One can do a similar decomposition for any $(J,I)$ pair and the results for a number of different pairs are shown in Table \ref{BreakItDown}. The decomposition is the same for the $S/T$ wavefunctions since the symmetry groups are isomorphic.

\bgroup
\def \arraystretch{1.2}
\begin{table}[!ht]
	\begin{center}
		\begin{tabular}{| c | c | } \hline
			$(J,I)$ or $(S,T)$ & Irreducible decomposition  \\ \hline
			$\left(\frac{1}{2},\frac{1}{2}\right)$ & $\boldsymbol{4}$ \\
			$\left(\frac{1}{2},\frac{3}{2}\right)$ & $\boldsymbol{2^+} \oplus \boldsymbol{2^-}\oplus\boldsymbol{4}$ \\
			$\left(\frac{3}{2},\frac{1}{2}\right)$ & $\boldsymbol{2^+} \oplus \boldsymbol{2^-}\oplus\boldsymbol{4}$ \\
			$\left(\frac{5}{2},\frac{1}{2}\right)$ & $\boldsymbol{2^+} \oplus \boldsymbol{2^-}\oplus\boldsymbol{4}\oplus\boldsymbol{4}$ \\
			$\left(\frac{7}{2},\frac{1}{2}\right)$ & $\boldsymbol{2^+} \oplus \boldsymbol{2^-}\oplus\boldsymbol{4}\oplus\boldsymbol{4}\oplus\boldsymbol{4}$ \\
			$\left(\frac{3}{2},\frac{3}{2}\right)$ & $\boldsymbol{2^+} \oplus \boldsymbol{2^-}\oplus \boldsymbol{4}\oplus\boldsymbol{4}\oplus\boldsymbol{4}$ \\ \hline
			
		\end{tabular}
		\vskip 7pt
		\caption{The decomposition of spin states into irreducible parts.}
		\label{BreakItDown}
	\end{center}
\end{table}
\egroup

Using Table \ref{BreakItDown} we can quickly see which combinations of $J,I,S$ and $T$ give permissible wavefunctions, and how many exist. A permissible wavefunction exists if there is a matching irreducible factor between the $J/I$ part and the $S/T$ part. For instance, there is one allowed state with $(J,I,S,T) = (1/2,1/2,1/2,1/2)$, since there is one factor of $\boldsymbol{4}$ in common for $(J,I) = (1/2,1/2)$ and $(S,T)=(1/2,1/2)$. There are four allowed states with $(J,I,S,T) = (3/2,1/2,5/2,1/2)$: two arising from the $\boldsymbol{4}$ irreps, one from the $\boldsymbol{2^+}$ irrep and one from the $\boldsymbol{2^-}$ irrep. There is only one allowed state with $(J,I,S,T) = (1/2,1/2,3/2,1/2)$.  We denote the $i$th wave function, whose $J/I$ and $S/T$ parts each transform as the irrep $\boldsymbol{X}$, having spins $(J, I, S, T)$ and space-fixed projection $t_3’$ as
\begin{equation}
\ket{J \, I \, S \, T; \boldsymbol{X_i} \, t_3'} \, .
\end{equation} 
We omit $i$ if there is only one such state.

To explicitly construct the permissible wavefunctions, we must choose a concrete realization of the irreps. Here, we make such a choice. The $\boldsymbol{4}$ irrep descends from the fundamental representation of $SU(2)\times SU(2)$. Hence an obvious choice of basis is the usual spin states\footnote{Note that the spin states $\Ket{J j_3 J}$ and $\Ket{I i_3 I}$ are not the usual spin states. Their generating matrices are conjugate to the usual ones.} with spins $(1/2, 1/2)$. The transformations then correspond to the matrices
\begin{equation}
N^{\boldsymbol{4}}_{C_3} = \frac{1}{\sqrt{2}}\begin{pmatrix} e^{\frac{i \pi}{12}} & 0 & e^{-\frac{5 i \pi}{12}} & 0 \\ 0 & e^{-\frac{7 i \pi}{12}} & 0 & e^{\frac{11 i \pi}{12}}   \\
e^{\frac{i \pi}{12}} & 0 & e^{\frac{7 i \pi}{12}} & 0 \\
0 & e^{-\frac{7 i \pi }{12}} & 0 & e^{-\frac{ i \pi}{12}}   \end{pmatrix} \text{ and } \, N^{\boldsymbol{4}}_{C_4} = \begin{pmatrix} 0 & e^{\frac{i \pi}{4}} & 0 &  0 \\  e^{\frac{ i \pi}{4}} & 0 & 0 & 0   \\
0 & 0 & 0 & e^{\frac{3 i \pi}{4}} \\
0 & 0 & e^{\frac{3 i \pi}{4}} & 0  \end{pmatrix}\, .\label{4trans}
\end{equation} 
The two-dimensional irreps are simpler. For the $\boldsymbol{2^+}$ irrep we may use a basis which transforms as
\begin{equation}
N^{\boldsymbol{2^+}}_{C_3} = \frac{1}{\sqrt{2}} \begin{pmatrix}
e^{-\frac{3 i \pi}{4}} & e^{\frac{3 i \pi}{4}} \\
e^{\frac{i \pi}{4}} & e^{\frac{3 i \pi}{4}} 
\end{pmatrix} \text{ and }\, N^{\boldsymbol{2^+}}_{C_4} = \frac{1}{\sqrt{2}} \begin{pmatrix}
e^{\frac{3 i \pi}{4}} & 0 \\
0 & e^{\frac{ i \pi}{4}} 
\end{pmatrix} \, . \label{2ptrans}
\end{equation}
Finally, there is a basis for the $\boldsymbol{2^-}$ irrep which transforms as
\begin{equation}
N^{\boldsymbol{2^-}}_{C_3} = \frac{1}{\sqrt{2}} \begin{pmatrix}
e^{-\frac{3 i \pi}{4}} & e^{\frac{3 i \pi}{4}} \\
e^{\frac{i \pi}{4}} & e^{\frac{3 i \pi}{4}} 
\end{pmatrix} \text{ and }\, N^{\boldsymbol{2^-}}_{C_4} = \frac{1}{\sqrt{2}} \begin{pmatrix}
e^{-\frac{ i \pi}{4}} & 0 \\
0 & e^{-\frac{ 3 i \pi}{4}} 
\end{pmatrix} \, . \label{2mtrans}
\end{equation}

We can now construct a wavefunction which satisfies \eqref{FRs2} and \eqref{FRs1} and we do so for $(J,I,S,T) = \left(1/2,1/2,1/2,1/2 \right)$. There is a set of $(J,I)=(1/2,1/2)$ states which transform as $\boldsymbol{4}$ and in particular transform as \eqref{4trans} under the action \eqref{JIsymm}. These are given by
\begin{align}
\Ket{\Psi_i}  = \big( \Ket{\tfrac{1}{2}\, \tfrac{1}{2}\, \tfrac{1}{2}} \Ket{\tfrac{1}{2}\, {-\tfrac{1}{2}}\, \tfrac{1}{2}}, \,  -\Ket{\tfrac{1}{2}\, \tfrac{1}{2}\, \tfrac{1}{2}} \Ket{\tfrac{1}{2}\, \tfrac{1}{2}\, \tfrac{1}{2}}, \,  \\ -i \Ket{\tfrac{1}{2}\, {-\tfrac{1}{2}}\, \tfrac{1}{2}} \Ket{\tfrac{1}{2}\, \tfrac{1}{2}\, \tfrac{1}{2}}, \,  i\Ket{\tfrac{1}{2}\, {-\tfrac{1}{2}}\, \tfrac{1}{2}} \Ket{\tfrac{1}{2}\, {-\tfrac{1}{2}}\, \tfrac{1}{2}}  \big)_i \,  . \label{halfbasis}
\end{align}
The $(S,T)=(1/2,1/2)$ states which transform as \eqref{phitrans} where $N$ is given by \eqref{4trans} are
\begin{align}
\Phi_i  = \big( \Ket{\tfrac{1}{2}\, \tfrac{1}{2}\, t_3'} \Ket{\tfrac{1}{2}\, {-\tfrac{1}{2}}\, {-t_3'}}, \,  -\Ket{\tfrac{1}{2}\, \tfrac{1}{2}\, t_3'} \Ket{\tfrac{1}{2}\, \tfrac{1}{2}\, {-t_3'}}, \,  \\ i \Ket{\tfrac{1}{2}\, {-\tfrac{1}{2}}\, t_3'} \Ket{\tfrac{1}{2}\, \tfrac{1}{2}\, {-t_3'}}, \,  -i\Ket{\tfrac{1}{2}\, {-\tfrac{1}{2}}\, t_3'} \Ket{\tfrac{1}{2}\, {-\tfrac{1}{2}}\, {-t_3'}}  \big)_i \,  .
\end{align}
 There are two sets of states which are identical apart from their space-fixed projections in $S/T$ space. Note that the bases $\ket{\Psi_i}$ and $\Phi_i$ are simply related -- their coefficients are complex conjugates. This is true in general since the $S/T$ wavefunctions must satisfy \eqref{phitrans}. Combining these two bases gives two wavefunctions with $(J,I,S,T)=(1/2,1/2,1/2,1/2)$ which are permitted by the cubic symmetry of the system. They are
\begin{align}  \Ket{\tfrac{1}{2} \, \tfrac{1}{2} \, \tfrac{1}{2} \, \tfrac{1}{2};\boldsymbol{4} \, t_3'} =\frac{1}{2}\Big(\Ket{\tfrac{1}{2}\, \tfrac{1}{2}} \Ket{\tfrac{1}{2}\, {-\tfrac{1}{2}}}\Ket{\tfrac{1}{2}\, \tfrac{1}{2}\, t_3'} \Ket{\tfrac{1}{2}\, {-\tfrac{1}{2}}\, {-t_3'}}+\Ket{\tfrac{1}{2}\, \tfrac{1}{2}} \Ket{\tfrac{1}{2}\, \tfrac{1}{2}}\Ket{\tfrac{1}{2}\, \tfrac{1}{2}\, t_3'} \Ket{\tfrac{1}{2}\, \tfrac{1}{2}\, {-t_3'}} + \nonumber \\\Ket{\tfrac{1}{2}\, {-\tfrac{1}{2}}} \Ket{\tfrac{1}{2}\, \tfrac{1}{2}}\Ket{\tfrac{1}{2}\, {-\tfrac{1}{2}}\, t_3'} \Ket{\tfrac{1}{2}\, \tfrac{1}{2}\, {-t_3'}}+\Ket{\tfrac{1}{2}\, {-\tfrac{1}{2}}} \Ket{\tfrac{1}{2}\, {-\tfrac{1}{2}}}\Ket{\tfrac{1}{2}\, {-\tfrac{1}{2}}\, t_3'} \Ket{\tfrac{1}{2}\, {-\tfrac{1}{2}}\, t_3'} \Big)\,  \label{gross}  
\end{align}
where we have suppressed the space-fixed projections of $J$ and $I$ for economy. To find wavefunctions with larger values of $J, I, S$ and $T$ we can simply repeat the process. The wavefunctions are large, complicated objects, as one would expect from quantization of an 11-dimensional system. We tabulate the bases of spin states which transform as \eqref{4trans}, \eqref{2ptrans} and \eqref{2mtrans} in Appendix \ref{app:A}. These can be used to construct wavefunctions with high spins.

\section{Parity and additional symmetries} \label{sec:5}

\subsection{Parity}

\begin{figure}[!ht]
	\centering
	\includegraphics[width=0.98\textwidth]{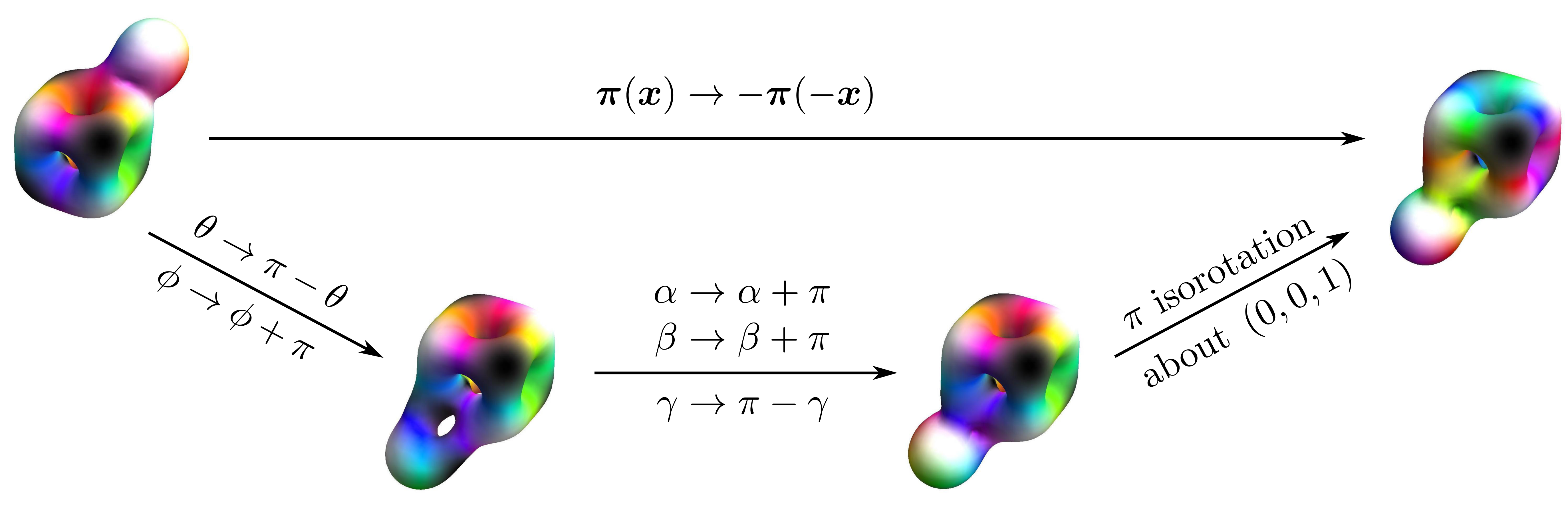} 
	\caption{ A visual representation of the equivalence between the inversion operator acting on the Skyrme field and its realization in our configuration space. } \label{parity}
\end{figure}

In the Skyrme model, the inversion operator is easily expressed in terms of the pion field, $\boldsymbol{\pi}(\boldsymbol{x})$. It is
\begin{equation}
\hat{P}: \boldsymbol{\pi}(\boldsymbol{x}) \to -\boldsymbol{\pi}(-\boldsymbol{x}) \, . \label{Pa}
\end{equation}
This inverts the orientation of the Skyrmion in space and in isospace. To apply the operator to our states we must first express it in terms of the coordinates on the configuration space. We can split this into two parts, one acting on the overall spin and isospin space $\hat{P}_{J,I}$ and another acting on the relative space $\hat{P}_{S,T}$. For our set-up $\hat{P}_{J,I}$ is easier to express using a momentum operator
\begin{equation}
\hat{P}_{J,I} = \text{exp}\left( -i \hat{I}_3 \pi \right) \, . \label{PJI}
\end{equation}
while the relative part is simplest to describe using an explicit coordinate transformation 
\begin{equation}
\hat{P}_{S,T} : (\theta,\phi,\alpha,\beta,\gamma) \to (\pi-\theta,\phi+\pi,\alpha+\pi,\beta+\pi,\pi-\gamma) \, . \label{PST}
\end{equation}
The equivalence between \eqref{Pa} and the combined action of \eqref{PJI} and \eqref{PST} is shown in Figure \ref{parity}, at a generic point in configuration space.

Using properties of the Wigner functions, we find that the inversion operator acts on the basis states as
\begin{equation}
\hat{P}\Ket{J\, j_3}\Ket{I \, i_3}\Ket{S\, s_3 \, {-t_3'}}\Ket{T\, t_3\, t_3'} = (-1)^{1+S+T+t_3-i_3}\Ket{J\, j_3 }\Ket{I \, i_3}\Ket{S\, s_3 \, t_3'}\Ket{T\, t_3\, {-t_3'}} \, .
\end{equation}
This can be used to construct states with definite parity $P$ which satisfy $\hat{P}\ket{\Psi} = P\ket{\Psi}=\pm\ket{\Psi}$. To find definite parity wavefunctions we begin with a set of degenerate energy eigenstates. We have not yet constructed such states but will do so soon. There are only two states with $J=I=S=T=1/2$, degenerate in their $t_3'$ value. The space-fixed projection cannot affect the energy and so these states, displayed in \eqref{gross}, must be energy eigenstates and can be used as a basis for the definite parity states. The definite parity states are
\begin{equation}
\ket{P = \pm 1} = \frac{1}{\sqrt{2}}\left(\Ket{\tfrac{1}{2} \, \tfrac{1}{2} \, \tfrac{1}{2} \, \tfrac{1}{2};\boldsymbol{4} \, \tfrac{1}{2}} \pm \Ket{\tfrac{1}{2} \, \tfrac{1}{2} \, \tfrac{1}{2} \, \tfrac{1}{2};\boldsymbol{4} \, {-\tfrac{1}{2}}}\right) \, . \label{defpar}
\end{equation}

\begin{figure}[!ht]
	\centering
	\includegraphics[width=0.9\textwidth]{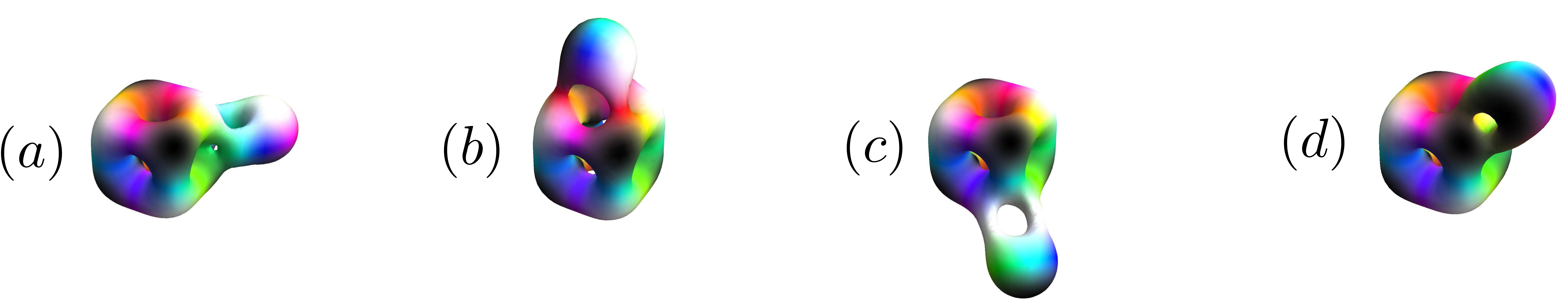} 
	\caption{Four points in the configuration space with a reflection symmetry. The $B=1$ lies at an edge, a face, a corner and a different edge of the cube in $(a)$, $(b)$, $(c)$ and $(d)$ respectively. } \label{Enhanced}
\end{figure}

There is an alternative way to calculate the parity at certain points in configuration space -- those where the Skyrme configuration has a reflection symmetry. At these, $\hat{P}$ can be written in terms of the $\hat{J}$ and $\hat{I}$ momentum operators. This gives a non-trivial consistency check on the global parity operator. As an example, consider the configuration displayed in Figure \ref{Enhanced}(a). This is the point $\boldsymbol{\alpha_z} = (\theta,\phi,\alpha,\beta,\gamma) = (\pi/2,\pi/4,0,\pi/2,0)$. At this point the parity operator is
\begin{equation}
\hat{P}\rvert_{\boldsymbol{\alpha_z}} = \exp\left(-i \pi \hat{J}_3 \right) \exp\left( i \pi \hat{I}_3 \right) \, . \label{parityz}
\end{equation}
We may apply this operator to the wavefunction evaluated at $\boldsymbol{\alpha_z}$ and should obtain the same parity as before. For the negative parity wavefunction in \eqref{defpar} we find that
\begin{align}
\hat{P}\rvert_{\boldsymbol{\alpha_z}}\Ket{ P=-1}\rvert_{\boldsymbol{\alpha_z}} &=\frac{1}{8\sqrt{2}} e^{-i \pi \hat{J}_3 + i \pi \hat{I}_3 }\Big( e^{\tfrac{i \pi}{8}}\Ket{\tfrac{1}{2} \, \tfrac{1}{2} }\Ket{\tfrac{1}{2} \, {-\tfrac{1}{2}} } +  e^{-\frac{i \pi}{8}}\Ket{\tfrac{1}{2} \, {-\tfrac{1}{2}} }\Ket{\tfrac{1}{2} \, \tfrac{1}{2} } \Big) \\
&= -\Ket{ P=-1}\rvert_{\boldsymbol{\alpha_z}}  \label{fullpar} \, ,
\end{align}
as expected. If the global parity operator is particularly difficult to write down, one could use this local method to construct definite parity states. For large spins, one would need to evaluate the wavefunction at several different points in configuration space where the configuration has a reflection symmetry. More of these are displayed in Figure \ref{Enhanced}.

Each energy eigenstate (which will be constructed in Section \ref{sec:6}) has a $2n$ degeneracy where
\begin{equation}
2n = \text{min}\left( 2S+1,2T+1 \right) \, .
\end{equation}
Half of these have positive parity and the other half have negative parity. Hence, there is at least parity doubling for each state. This is not seen in the $^5$He/$^5$Li energy spectra. The doubling is ultimately due to the lack of symmetry in our configuration space. The configurations with the largest symmetry group include those seen in Figure \ref{Enhanced}(a) and (d) which have an overall $C_2$ symmetry. This order $2$ group partitions the total spin/isospin basis into two sets -- those which transform trivially under the $C_2$ element and those which pick up a sign. One of these sets is disregarded due to the FR constraints. The remaining, allowed, wavefunctions are once again partitioned in two by the order $2$ parity operator. For half-integer spin and isospin, the set of free states is divisible by four and so each positive parity state has a negative parity partner. For example there are four $J=I=1/2$ basis states. Two are allowed by the FR constraints: one has positive parity and the other has negative parity. By this reasoning, if a Skyrmion has only $C_2$ symmetry, its energy spectrum will contain approximate parity doubling\footnote{The structure of the moment of inertia tensor may break the degeneracy slightly.}, almost never seen in experimental data. This appears to be a problem for Skyrme models with low classical binding energy. For example, the lightly bound model \cite{GHKMS} contains parity doubling for many of its Skyrmions. This argument suggests that one should attempt to construct a Skyrme model with low classical binding energy and high symmetry. There have been several attempts to write down such a model. One is to include vector mesons in the Lagrangian. Recently it was shown that inclusion of the rho meson reduces the classical binding energy by two-thirds, while retaining the symmetry group of each Skyrmion up to baryon number eight \cite{SN}. If one integrates out the mesons, a term which contains sixth order derivatives of the pion field appears \cite{AN}. If one only includes this term, the theory is BPS (i.e. has no classical binding energy) and so, unsurprisingly, its inclusion lowers the binding energy \cite{ASW}. Numerical work \cite{GHS} has shown that the classical symmetries do remain for a family of Lagrangians where the sixth order term does not completely dominate. One can also modify the pion potential term which, again, has been shown to reduce binding energies while retaining much of the Skyrmion symmetries \cite{Gud,Gud2,Gud3}. In all of these models, quantum effects arising from the spin of the $B=1$ Skyrmion spoil the small classical binding energies. An alternative approach is to start with a classically tightly bound but highly symmetric Skyrme model and reduce the binding energy by taking account of more modes in the quantization procedure, as suggested in \cite{Hal,BT}. If the classical $B=5$ Skyrmion has a symmetry group of order 4 the parity doubling problem is avoided. 

Finally, we compare our results about parity with the shell model, which does not contain doubling. The key difference is that, in the shell model, the core and additional particle are individual \emph{quantum} objects. They are then combined to make a two-cluster system. Since the core is already quantized it has spherical symmetry and the additional particle is governed by a central potential. In our case, the additional particle feels the classical structure of the core, which has much less symmetry than the spherical core of the shell model. This reduction in symmetry means that relatively more states are allowed in our model, leading to incorrect results. Our clusters are combined at the classical level and then quantized. We could perhaps reproduce the shell model results by quantizing the $B=4$ and $B=1$ separately and then combining them. It may be of interest to compare these two methods and find out when each is appropriate.

\subsection{$C_2$ symmetry}

The configuration space appears to contain more symmetries than
we have considered so far. Here, we will consider one such symmetry
and explain why it is in fact included in our calculation. 

Consider the point in Figure \ref{Enhanced}(a), $\boldsymbol{\alpha_z}$. This has a $C_2$ symmetry, realized as
\begin{equation}
\exp\left(-\frac{\pi i }{\sqrt{2}}\left(\hat{J}_1+\hat{J}_2\right) - \pi i \hat{I}_1\right) \Ket{\Psi}\rvert_{\boldsymbol{\alpha_z}} = \Ket{\Psi}\rvert_{\boldsymbol{\alpha_z}} \, . \label{C2atpt}
\end{equation}
The cubic group that we considered earlier includes a closely related $C_2$ element
\begin{equation}
\exp\left(-\frac{\pi i }{\sqrt{2}}\left(\hat{J}_1+\hat{J}_2\right) - \pi i \hat{I}_1+\frac{\pi i }{\sqrt{2}}\left(\hat{S}_1+\hat{S}_2\right) + \pi i \hat{T}_1\right) \Ket{\Psi} = \Ket{\Psi} \, . \label{fullC2}
\end{equation}
The $S/T$ part of this transformation can be viewed as a coordinate transformation on the $S/T$ space. It takes $\boldsymbol{\alpha_z}$ to itself. Hence
\begin{equation}
\left(\frac{\pi i }{\sqrt{2}}\left(\hat{S}_1+\hat{S}_2\right) + \pi i \hat{T}_1\right)\Ket{\Psi}\rvert_{\boldsymbol{\alpha_z}} = \Ket{\Psi}\rvert_{\boldsymbol{\alpha_z}} \,
\end{equation}
meaning that the $S/T$ part of \eqref{fullC2} is trivial at the point $\boldsymbol{\alpha_z}$ and thus \eqref{C2atpt} is satisfied.

Provided that the constraint \eqref{C2atpt} is satisfied, several other constraints are automatically dealt with. Denote the point in Figure \ref{Enhanced}(d) as $\boldsymbol{\alpha_y} = (\pi/4,0,4\pi/3,\pi / 2,\pi /2)$. This should satisfy the constraint
\begin{equation}
\hat{C}_2^{\boldsymbol{y}}\Ket{\Psi}\rvert_{\boldsymbol{\alpha_y}} \equiv \text{exp}\left(-\frac{\pi i}{\sqrt{2}} \left(\hat{J}_1+\hat{J}_3 \right)  -  \pi i\left( \cos(4\pi/3)\hat{I}_1+\sin(4\pi/3)\hat{I}_2\right) \right)\Ket{\Psi}\rvert_{\boldsymbol{\alpha_y}} = \Ket{\Psi}\rvert_{\boldsymbol{\alpha_y}} \, .
\end{equation}
To show that this does hold, we will need to introduce some extra terminology. Denote the $C_3$ operator which relates the configurations in Figure \ref{Enhanced}(a) and Figure \ref{Enhanced}(d) in $S/T$ space as
\begin{equation}
\hat{C}^{S,T}_3=\text{exp}\left( \frac{2\pi i}{3 \sqrt{3}} \left(\hat{S}_1+\hat{S}_2+\hat{S}_3\right)  +  \frac{2\pi i}{3} \hat{T}_3\right) \, .  \label{C3yis}
\end{equation} 
The total wavefunction has been constructed to be invariant under \eqref{C3yis} combined with the related action on the $J/I$ space
\begin{equation}
\hat{C}^{J,I}_3=\text{exp}\left(- \frac{2\pi i}{3 \sqrt{3}} \left(\hat{J}_1+\hat{J}_2+\hat{J}_3\right)  -  \frac{2\pi i}{3} \hat{I}_3\right) \, .
\end{equation}
Hence the total wavefunction satisfies
\begin{equation}
 \hat{C}_3^{S,T} \hat{C}_3^{J,I} \Ket{\Psi} = \Ket{\Psi} \, , \label{const}
\end{equation}
which means that
\begin{equation}
\Ket{\Psi}\rvert_{\boldsymbol{\alpha_z}}  = \left(\hat{C}_3^{S,T}\right)^{-1}\Ket{\Psi}\rvert_{\boldsymbol{\alpha_y}} = \hat{C}_3^{J,I} \Ket{\Psi}\rvert_{\boldsymbol{\alpha_y}} \, .
\end{equation}
The first equality is simply a coordinate transformation on the $S/T$ space while the second is a consequence of \eqref{const}.

Now we can consider the wavefunction at $\boldsymbol{\alpha_y}$. It satisfies
\begin{align}
\hat{C}_2^{\boldsymbol{y}}\Ket{\Psi}\rvert_{\boldsymbol{\alpha_y}} &= \left(\hat{C}^{J,I}_3\right)^{-1}\left(\hat{C}^{J,I}_3\hat{C}_2^{\boldsymbol{y}} \left(\hat{C}^{J,I}_3\right)^{-1}\right)\left( \hat{C}^{J,I}_3\Ket{\Psi}\rvert_{\boldsymbol{\alpha_y}}\right)\non
&= \left(\hat{C}^{J,I}_3\right)^{-1}\hat{C}_2^{\boldsymbol{z}}\Ket{\Psi}\rvert_{\boldsymbol{\alpha_z}}   \non
&= \left(\hat{C}^{J,I}_3\right)^{-1} \Ket{\Psi}\rvert_{\boldsymbol{\alpha_z}} \non
&= \Ket{\Psi}_{\boldsymbol{\alpha_y}} \, .
\end{align}
Hence the wavefunction obeys the constraint at $\boldsymbol{\alpha_y}$ provided it also obeys the constraint at $\boldsymbol{\alpha_z}$, due to the group structure. This type of argument is rather straightforward but it contains a key message: get the symmetry of the configuration space correct and everything else will follow.

\subsection{Enhanced $D_2$ symmetry}

\begin{figure}[!ht]
	\centering
	\includegraphics[width=0.9\textwidth]{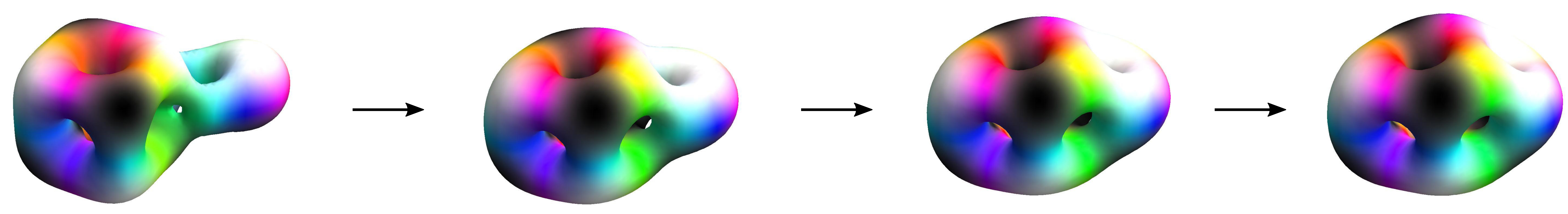} 
	\caption{Gradient flow from the configuration in Figure \ref{Enhanced}(a) to the $D_2$-symmetric Skyrmion. This process is carried out in the standard Skyrme model with dimensionless pion mass $m_1=1$.} \label{D2flow}
\end{figure}

There is an additional symmetry we may consider. This exists just outside of our configuration space. Consider the point $\boldsymbol{\alpha_z}$ in Figure \ref{Enhanced}(a). In the standard Skyrme model, when the $B=1$ Skyrmion is brought closer to the cube, the configuration deforms into the well known $D_2$-symmetric $B=5$ Skyrmion. This deformation process is displayed in Figure \ref{D2flow}. The $B=5$ Skyrmion has $D_2$ symmetry, generated by two elements. The first is the $C_2$ symmetry \eqref{fullC2} and the other is another $C_2$ element which enforces an extra constraint on the wavefunction at the point $\boldsymbol{\alpha_z}$
\begin{equation}
\text{exp}\left(- i \pi  \left(\tfrac{1}{2} \hat{J}_1-\tfrac{1}{2}\hat{J}_2+\tfrac{1}{\sqrt{2}}\hat{J}_3\right) - i \pi \left(\tfrac{1}{\sqrt{2}} \hat{I}_2-\tfrac{1}{\sqrt{2}} \hat{I}_3   \right) \right)\Ket{\Psi}\rvert_{\boldsymbol{\alpha_z}} = -\Ket{\Psi}\rvert_{\boldsymbol{\alpha_z}} \, . \label{D4FR}
\end{equation}
We calculated the FR-sign using the approach of Krusch based on the rational map approximation \cite{Kru}. We can check if our definite parity energy eigenstates obey the constraints by evaluating the wavefunctions at $\boldsymbol{\alpha_z}$. When there are degenerate eigenstates ($S,T>1/2$) we can attempt to construct linear combinations of the states which obey the constraints. If the wavefunction satisfies the constraint at $\boldsymbol{\alpha_z}$ it is automatically enforced at the other $D_2$-symmetric points due to the argument used in the previous subsection.

We do not rule out states which do not satisfy \eqref{D4FR} since our configuration space does not explicitly include the $D_2$-symmetric Skyrmions. However their existence still affects the energies of our states. To see why, consider a quantization which includes a parameter $\tau$ describing the flow seen in Figure \ref{D2flow} (it could be thought of as the radial coordinate) and suppose the $D_2$ symmetry is restored at $\tau^*$. The wavefunctions of this system will look schematically like
\begin{equation}
u(\tau)\times \Ket{\Psi} \, ,
\end{equation}
where the $\ket{\Psi}$ are the wavefunctions we have calculated in this paper. Now, consider a ``rotational" state which is allowed by the $D_2$ symmetry, denoted $\ket{\Psi;D}$. Here the wavefunction $u(\tau)$ coupled to $\ket{\Psi;D}$ is not restricted in any way. Now consider a rotational wavefunction not allowed by the $D_2$ symmetry \eqref{D4FR}, denoted $\ket{\Psi;C}$. If this is coupled to a wavefunction $u(\tau)$, then the only way to satisfy the $D_2$ symmetry \eqref{D4FR} is for $u(\tau)$ to vanish at $\tau^*$. This imposes an extra constraint on the wavefunction. Constrained wavefunctions naturally have higher energy and so the total wavefunction containing $\ket{\Psi;C}$ has more energy than the one containing $\ket{\Psi;D}$. The implication for our model is that wavefunctions not allowed by \eqref{D4FR} should gain an additional energy contribution. We call this the constraint energy. If one thinks of $\tau$ as a radial coordinate, the $\ket{\Psi;D}$ states are coupled to a ground state radial wavefunction while the  $\ket{\Psi;C}$ states would be coupled to an excited radial wavefunction, since it must vanish at the minimum of the potential energy (for the standard Skyrme model).

The size of the constraint energy is hard to measure. To do so properly, one should consider a configuration space and corresponding quantization scheme which includes a coordinate like $\tau$. Even to estimate the contribution, one needs to understand the distance between the configurations we consider in our model and the $D_2$-symmetric Skyrmion. This depends crucially on the interpretation of our configuration space $\mathcal{M}$, briefly mentioned in Section \ref{sec:2}. One may interpret $\mathcal{M}$ as already including the $D_2$ Skyrmion at the point $\boldsymbol{\alpha_z}$. The relative coordinates between the clusters are then internal, vibrational excitations of the $D_2$-symmetric $B=5$ Skyrmion. The two-cluster picture is then just a convenient way to think about these coordinates. In this picture, the constraint energy is very large. To see if this interpretation is possible, one would need to do a small amplitude analysis of the $B=5$ Skyrmion to see if there is a map between its vibrational degrees of freedom and those in this model. The true picture is likely somewhere between this interpretation and the two-weakly-bound-clusters interpretation. To understand it properly, a more thorough investigation of the $B=5$ configuration space is required.

\section{Calculating Energies} \label{sec:6}

\subsection{Kinetic Energy}

The Hamiltonian $\mathcal{H}$ of the system is simple to express in terms of the body-fixed momenta of the individual Skyrmions. Ignoring deformations caused by the Skyrmion interaction which will change the moment of inertia tensors, the kinetic part of the energy is
\begin{equation}
\mathcal{H}_{\text{free}} = \frac{1}{2}\left( \boldsymbol{L}^{(4) T}V_4^{-1}\boldsymbol{L}^{(4)} + \boldsymbol{K}^{(4) T}U_4^{-1}\boldsymbol{K}^{(4)} +\boldsymbol{L}^{(1) T}V_1^{-1}\boldsymbol{L}^{(1)} +(M_1R^2)^{-1} \, \boldsymbol{l}^T\boldsymbol{l} \right) \, ,
\end{equation}
where $V_{i}$ and $U_{i}$ are the moment of inertia tensors associated with the angular and isoangular motion of the $B=i$ Skyrmion, $M_1$ is the mass of the $B=1$ Skyrmion and $R=| \boldsymbol{R}|$ is the (fixed) distance between the Skyrmions. We can express the Hamiltonian in terms of the new momenta by inverting equations \eqref{Jdef}, \eqref{Idef} and \eqref{STdef}. By doing this we find that
\begin{equation}
\begin{pmatrix} \boldsymbol{L}^{(4)} \\ \boldsymbol{K}^{(4)} \\ \boldsymbol{L}^{(1)} \\ \boldsymbol{l} \end{pmatrix} = \begin{pmatrix} \boldsymbol{1} & 0 & -\boldsymbol{1} & 0 \\ 0 & \boldsymbol{1} & 0 & -\boldsymbol{1} \\ 0 & 0 & 0 & -B^{(1)T}B^{(4)} \\ 0 & 0 & A^{(4)} & A^{(1)}B^{(1)T}B^{(4)} \end{pmatrix}\begin{pmatrix}
\boldsymbol{J} \\ \boldsymbol{I} \\ \boldsymbol{S} \\ \boldsymbol{T}
\end{pmatrix} \, .
\end{equation}
where $\boldsymbol{1}$ represents the $3\times3$ unit matrix.  The Hamiltonian becomes
\begin{align}
  &\mathcal{H}_{\text{free}} =
  \frac{1}{2}\begin{pmatrix}
  \boldsymbol{J}&\boldsymbol{I}&\boldsymbol{S}&\boldsymbol{T}\end{pmatrix}
  \;\mathcal{G}\, \begin{pmatrix}
  \boldsymbol{J} \\ \boldsymbol{I} \\ \boldsymbol{S} \\ \boldsymbol{T}
  \end{pmatrix} \, , \label{kin}\\
  &\mathcal{G}\equiv
  \begin{pmatrix}
V_4^{-1} & 0 & -V_4^{-1} & 0 \\0 & U_4^{-1} & 0 & -U_4^{-1} \\ -V_4^{-1} & 0 & V_4^{-1} + (M_1R^2)^{-1} & -(M_1R^2)^{-1} A^{(4)T} A^{(1)} B^{(1)T } B^{(4)} \\ 0 & -U_4^{-1} &  -(M_1R^2)^{-1} A^{(4)T} A^{(1)} B^{(1)T} B^{(4)} & U_4^{-1} + V_1^{-1} + (M_1R^2)^{-1}
\end{pmatrix}\!.\nonumber
\end{align}
The only complicated term in this expression is the $S/T$ cross-term. We are unsure how to evaluate this term and so we replace it by its expectation value
\begin{equation}
(M_1 R^2)^{-1} A^{(4)T} A^{(1)} B^{(1)T} B^{(4)} \to (M_1 R^2)^{-1}\big<A^{(4)T} A^{(1)} B^{(1)T}  B^{(4)} \big> = 0 \, .
\end{equation}
Although we would ideally evaluate this term properly, we are comforted by the fact that $M_1R^2$ is the largest scale in the metric.

The moment of inertia tensors are all diagonal, and most are proportional to the unit matrix. We use small letters to describe the diagonal elements so that
\begin{equation}
V_4 = v_4 \boldsymbol{1}\, , \quad V_1 = v_1 \boldsymbol{1}\, \quad \text{and} \quad R^2 M_1 = \mu \boldsymbol{1} \, .
\end{equation}
The $B=4$ isospin tensor is slightly more complicated. There are two independent diagonal components. In the orientation we have used,
\begin{equation}
(U_4)_{11} = (U_4)_{22} = u_{11} \quad \text{and} \quad (U_4)_{33} = u_{33} \, .
\end{equation}
We fix the free moments of inertia and masses numerically using the results from the standard Skyrme model with dimensionless pion mass $m_1=1$. In Skyrme units, this gives
\begin{align}
v_4 = 661, \quad   u_{11} = 147, \quad u_{33} = 176, \quad M_4 = 613, \quad v_1 = 48 \quad \text{and} \quad M_1 = 168 \, .
\end{align}
We are left to fit $R$ and hence find $\mu$. We take $R=3$, and this length is also used when we numerically generate the interaction potential.

To find the energy eigenstates we simply diagonalize the Hamiltonian matrix for each set of allowed states with a fixed set of spins. We review the low lying states here and an extensive table of higher energy states can be found in Appendix \ref{app:B}. We list many more states than are experimentally seen. This is not because we are directly interested in them. Rather, we are interested in how their existence affects the lowest energy states, once the potential is turned on.

The ground state of the free system is the state calculated in \eqref{gross}
\begin{equation}
\Ket{\tfrac{1}{2}\,\tfrac{1}{2}\,\tfrac{1}{2}\,\tfrac{1}{2}; \boldsymbol{4} \, t_3'} \, .
\end{equation}
This has energy
\begin{equation}
E = \frac{3}{8v_1} + \frac{3}{4\mu} \, .
\end{equation}
The state is doubly degenerate, due to the free $t_3'$ label. There is one state with negative parity, permitted by the $D_2$-symmetric configuration, and one with positive parity which is not.  This state is simple to interpret: since there is no $V^{(4)}$ or $U^{(4)}$ dependence, the $B=4$ core is at rest while the orbiting nucleon has spin 1/2. Not all states have such a simple interpretation.

Just above the ground state there is a spin 3/2 state
\begin{equation}
\sqrt{2}\Ket{\tfrac{3}{2}\,  \tfrac{1}{2} \,\tfrac{3}{2} \,\tfrac{1}{2};\boldsymbol{4}\, t_3'} + \Ket{\tfrac{3}{2}  \tfrac{1}{2} \tfrac{3}{2} \tfrac{1}{2};\boldsymbol{2^+}\, t_3'}+\Ket{\tfrac{3}{2}  \tfrac{1}{2} \tfrac{3}{2} \tfrac{1}{2};\boldsymbol{2^-}\, t_3'}\,,
\end{equation}
which has energy
\begin{equation}
E = \frac{3}{8v_1} + \frac{9}{4\mu} \, .
\end{equation}
Once again, this is doubly degenerate with a negative parity state, permitted by the $D_2$ Skyrmion, and a positive parity state which is not. Of the four states discussed so far, the two negative parity states are identified with the two low-energy states of the $^5$He/$^5$Li isodoublet. Experimentally, there are no low-energy positive-parity states. Hence, to match data, the constraint energy from the $D_2$-symmetric Skyrmion must be large. This provides some evidence that the correct Skyrme Lagrangian should contain a low-energy $D_2$-symmetric $B=5$ Skyrmion. This is not true for lightly bound models such as \cite{Gud,Gud2,Gud3} and \cite{GHKMS}. It is unclear what happens near the BPS limit of the sextic model \cite{GHS}.

The experimental data then has a large gap of around $15$ MeV. Our spectrum also has a large gap. The next six states are energetically degenerate. They are
\begin{gather}
\Ket{\tfrac{3}{2}\,  \tfrac{1}{2}\, \tfrac{1}{2}\, \tfrac{1}{2};\boldsymbol{4}\, t_3'}\, ,  \\ 
\Ket{\tfrac{5}{2}\,  \tfrac{1}{2}\, \tfrac{1}{2}\, \tfrac{1}{2};\boldsymbol{4_1} \, t_3'}\, , \\
\text{and }\Ket{\tfrac{3}{2}\,  \tfrac{3}{2}\, \tfrac{1}{2}\, \tfrac{1}{2};\boldsymbol{4_2}\, t_3'} - \sqrt{3}\Ket{\tfrac{3}{2} \, \tfrac{3}{2} \,\tfrac{1}{2} \,\tfrac{1}{2};\boldsymbol{4_3} \, t_3'} \, ,
\end{gather}
where $t_3' = \pm \frac{1}{2}$. Each of these has a positive and negative parity version. They all have energy
\begin{equation}
E= \frac{3}{8v_1} +\frac{1}{2u_{11}}+\frac{1}{2u_{33}}+\frac{3}{v_4} +\frac{3}{4\mu} \, . \label{gaped}
\end{equation}
The large degeneracy in the spectrum may be expected since the model is a free theory with a simple kinetic operator. In the full model, the Skyrmions interact which alters the moment of inertia tensor. This will likely break the degeneracy of these states.

The energy spectrum then becomes rather dense -- there are many states with similar energies. Many of these can be found in Appendix \ref{app:B} and we plot their spectrum in Figure \ref{energyspec}. The main success of the free theory is the large energy gap in the spectrum.
 
\subsection{Potential Energy}

Having found the wavefunctions for the free system, we can now estimate the potential energy contribution. To find the potential $V(\theta,\phi,\alpha,\beta,\gamma)$ we insert a symmetrized product ansatz of a $B=1$ and $B=4$ Skyrmion into the static Skyrme Lagrangian. Numerically, we discretize the angles with a lattice spacing of $\pi/12$. This amounts to finding $V$ at approximately two million points.

The full Hamiltonian $\mathcal{H}$ is the kinetic operator \eqref{kin} plus the numerically generated potential. Denoting our free wavefunctions as $\Ket{\Psi^i}$ the energy spectrum is found by diagonalizing
\begin{equation}
\Bra{\Psi^i} V(\theta,\phi,\alpha,\beta,\gamma) \Ket{\Psi^j} \, .
\end{equation}
Although the wavefunctions depend on eleven coordinates, the part of the matrix element which depends on $J$ and $I$ is rather trivial. Hence to calculate the matrix element we only need to do an integration over the 5-dimensional $S/T$ space. States with different spins, isospins and parities have zero mixing and so we focus on one sector at a time. We only calculate a finite number of free energy eigenfunctions and so we diagonalize with respect to a truncated basis. We include all states that we found up to and including an energy cut-off, which is
\begin{equation}
E= \frac{15}{8v_1}+\frac{5}{2u_{11}}+\frac{1}{2u_{33}}+\frac{3}{v_4}+\frac{15}{4\mu} \, .
\end{equation}
This means that, for example, we include fourteen states with $(J,I)=(1/2,1/2)$ and twenty-seven states with $(J,I)=(3/2,1/2)$. We include less basis states when considering larger spins. Fortunately, we are primarily interested in the two low-spin cases, where we have already calculated a large basis of states.

\begin{figure}[!ht]
	\centering
	\includegraphics[width=0.98\textwidth]{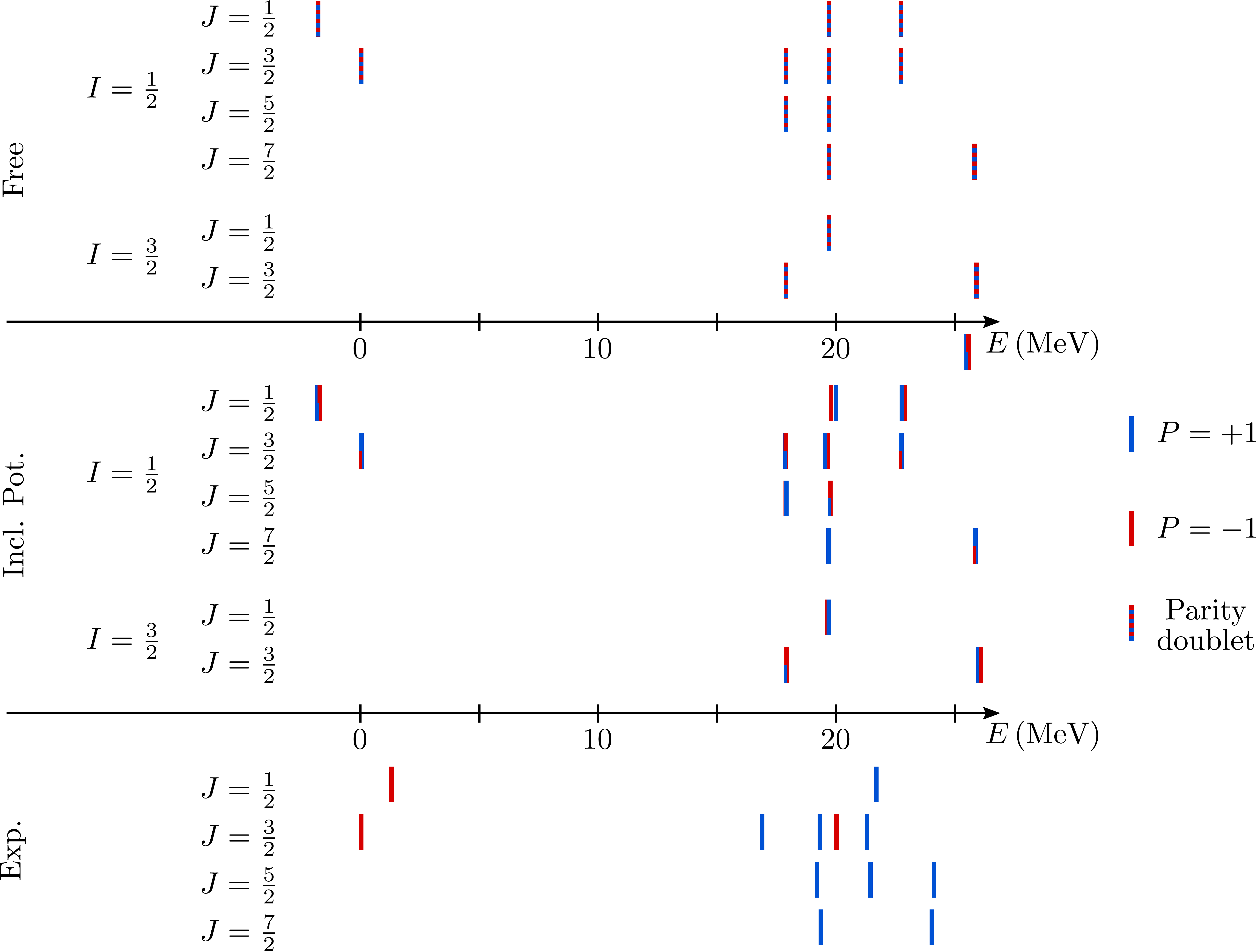} 
	\caption{The energy spectrum for the free theory and after the inclusion of the potential term, compared to the energy spectrum of $^5$He, taken from \cite{Til}. Bars which are red (blue) represent states with negative (positive) parity, while a striped bar represents a parity doublet --  two degenerate states with opposite parities. } \label{energyspec}
\end{figure}

The results of the calculation are shown in Figure \ref{energyspec}
alongside the free energy spectrum and the experimental data, which are taken from \cite{Til}. To calibrate the model we must choose energy and length scales. Or equivalently, choose an energy scale and the value of $\hbar$ in Skyrme units. We take $\hbar= 21.26$, the same value taken in \cite{MMW}. We use an energy scale where one Skyrme energy unit is $4.04$ MeV, this is $70\%$ smaller than the scale used in \cite{MMW}. Our calibration is different since we are using a different quantization scheme. However, it is encouraging that the two sets of parameters are reasonably similar.

The results contain some successes -- all experimental states are seen and the spectrum contains a large gap. There are more free spin $3/2$ states than $1/2$ states and so one may expect, with a larger basis to mix, the low lying spin $3/2$ and $1/2$ states would reverse their order. However, due to the details of the mixing, the spin $1/2$ states remain the lowest energy ones. Hence we do not get the correct ground state spin. The potential only has a small effect on the spectrum since the Skyrmions interact weakly where they are positioned. If we artificially increase the size of the potential, the states remain in the incorrect order. In fact, the size of the gap between the low energy spin $1/2$ and $3/2$ states \emph{increases}. This is the opposite of what one would expect from spin-orbit coupling. Hence the structure of the pion field may not account of the spin-orbit effect in the Skyrme model, at least at weak coupling. We calculated the potential for the Skyrme model with $m_2=0$ and $m_2=2.8$. The change of potential had little impact on the results.

The model also contains approximate parity doubling -- not seen in the $^5$He/$^5$Li spectra. We have already suggested one way to remedy the problem: by including the $D_2$-symmetric Skyrmion in the configuration space, or energetically punishing the states which are not permitted by this symmetry. This would add a constraint energy contribution to both the low lying $3/2^+$ and $1/2^+$ states, which are not permitted by this symmetry. In addition, we fail to obtain the $1/2^+$ ground state of the isospin $3/2$ nuclei, Hydrogen-$5$ and Beryllium-$5$, although there is a low energy $1/2^+$ state in our spectrum. These nuclei are highly unstable so our bound two-cluster model is likely unsuitable to accurately describe such states.

\section{Further Work} \label{sec:7}

The framework developed in this paper may be applied to a wide range of systems. For any strongly bound Skyrmion with baryon number $B$, our approach can be used to study the $B+1$ nucleus. In the Skyrme model, like many nuclear models, nuclei with $B=4N$ are particularly stable. For example, the $B=32$ Skyrmion is strongly bound and has cubic symmetry \cite{FLM}. Hence the $B=33$ Skyrmion's configuration space will likely have a low energy subspace which looks like a $32+1$ two-cluster system. This will model certain states of the Sulfur-33/Chlorine-33 isodoublet. Experimentally, the first two states of these nuclei have spin-parity $3/2^+$ and $1/2^+$ respectively. Applying the free model developed in this paper (which is rather naive), we would find these low lying states, alongside their negative parity partners. A more careful study may explain why the positive parity states are preferred.

There is evidence of large-radius ``Hoyle-like" states in the Carbon-13 spectrum \cite{ODB}. In the Skyrme model, these would be described by a single nucleon orbiting the $B=12$ chain-like Skyrmion, which models the Hoyle state. One could describe these novel states using the framework from this paper but rather than restricting the $B=1$ to a sphere, it should be restricted to an ellipse. In fact, one could restrict the $B=1$ to any surface which reflects the symmetry of the core Skyrmion, such as those with tetrahedral, octahedral or dodecahedral symmetry. This technique could be used to study nuclei with one nucleon more than the ``magic" tetrahedral Skyrmions discussed recently in \cite{Man} and \cite{HMR}. One could even repeat the calculation in this paper but insist that the $B=1$ is restricted to a cube, so that it is always just touching the $B=4$ core. Then the Wigner functions in the $S/T$ space used as a basis for the wavefunctions would be replaced with free wavefunctions on the surface of the cube.

In the lightly bound Skyrme model \cite{GHS,GHKMS}, large Skyrmions
are approximately described by a set of individual $B=1$ Skyrmions
which take positions on a face centered cubic (FCC) lattice. Here,
there are strongly bound Skyrmions which arise when a layer of the FCC
lattice is filled. The Skyrmion with one extra baryon is then
described by a core $+$ particle system. To quantize these, Manton
suggested that one should allow the extra Skyrmion to only take
positions on the next layer of the FCC lattice \cite{Man}. One must
still quantize the overall spin and isospin of the system and our
framework shows how to do so. It amounts to replacing the classical
configuration space 
\begin{equation}
SO(3)_J \times SO(3)_I \times \frac{SO(3)_S \times SO(3)_T}{U(1)}
\end{equation}
with 
\begin{equation}
SO(3)_J \times SO(3)_I \times C_n \, ,
\end{equation}
where $C_n$ is a collection of $n$ positions that the additional Skyrmion may take: those on the next layer of the FCC. The Hamiltonian is then a hopping Hamiltonian and the relative part of the wavefunction is a function on a finite set of points. The overall wavefunctions must still obey the FR constraints discussed in this paper but now the symmetry transformations are members of the symmetric group of $n$ points, rather than rotations in $SO(3)_S \times SO(3)_T$. This reduces the complexity of the problem as there is no longer a need to generate a potential on the relative space and the Schr\"odinger equation is likely exactly solvable. With this simplicity, one may then try to study more complex systems such as those with more than one orbiting nucleon. These include most of the halo nuclei and those with a few more nucleons than a magic nucleus. This quantization may even be relevant for the standard Skyrme model. For example, in the $B=5$ sector, we could restrict the $B=1$ to only take positions at the minima of the interaction potential between the $B=1$ and $B=4$ clusters. In our case, this is at the faces, edges and corners of the cube (with the $B=1$ orientated in the attractive channel). This will correspond to the tight binding limit rather than the weakly interacting limit we have considered in this paper. It would be interesting to compare the resulting spectra in each limit.

\begin{figure}[!ht]
	\centering
	\includegraphics[width=0.96\textwidth]{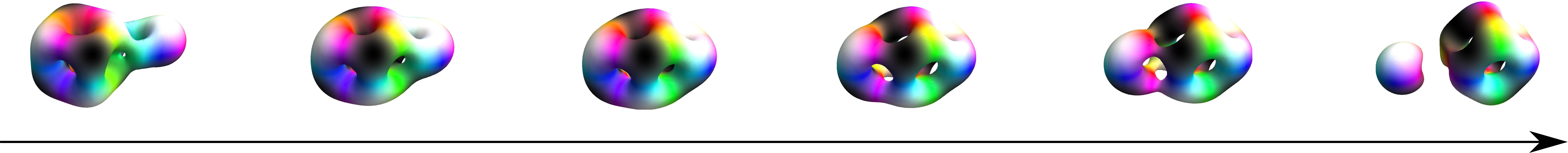} 
	\caption{A vibrational mode of the $4+1$ system.} \label{Scattering}
\end{figure}

The initial results of this paper are not very promising, but the model
can be improved in many ways. To match data, we will need to rely on the
existence of the $D_2$-symmetric $B=5$ Skyrmion. However, our
configuration space does not include it. We were able to find free
wavefunctions which do satisfy the constraint arising from the $D_2$-symmetric Skyrmion. However, after inclusion of the potential energy, these were
mixed with states which do not. Hence the overall state only
approximately satisfies the constraint. This inconsistency is
ultimately due to our exclusion of the $D_2$ configuration. We should
take account of the mode seen in Figure \ref{D2flow}. However, this
evolution is only half of a vibrational mode. The full mode is seen in
Figure \ref{Scattering}. A $B=1$ Skyrmion is positioned at an edge of
the $B=4$ cube. It merges with the cube to become the $D_2$ Skyrmion
but as this process continues, a different $B=1$ Skyrmion is knocked
out of the system. The original $B=1$ has become part of a leftover
cube, which is rotated and isorotated relative to the original cube. 

This novel vibrational mode, which creates a non-trivial relationship between the initial and final configurations of Figure \ref{Scattering}, should be included in the configuration space of the full model. Our calculation knows nothing of the non-trivial relationship and hence we cannot consistently create non-free wavefunctions which are permitted by the $D_2$ symmetry. The work of this paper will serve as an essential initial step to understanding the full model. In particular, the wavefunctions calculated here can be used as part of the eigenfunction basis for the more complicated system. Before progress can be made, a careful study of the $B=5$ vibrational space must be completed.

\section{Conclusion} \label{sec:8}

We have attempted to describe the $^5$He/$^5$Li isodoublet in the Skyrme model by treating the $B=5$ Skyrmion as a two-cluster system. To do so, we found coordinates where the overall rotations and isorotations are factored out of the full configuration space. This step is essential for any core + particle system in the Skyrme model. It also drastically reduces the complexity of the problem since the interaction potential only depends on the relative coordinates between the clusters; a 5-dimensional space rather than the original 11-dimensional space.

We then constructed wavefunctions permitted by the cubic symmetry of the $B=4$ core. This construction relied on the representation theory of the system's constituent symmetry groups. The same construction may be applied to any system where one considers a configuration space of deformed Skyrmions. The representation theory was not enough to construct energy eigenfunctions of definite parity states and standard techniques were used to determine these.

Until this point the work was rather model independent, relying only on the cubic symmetry of the $B=4$ core. To calculate the energy spectrum, we had to choose a specific Skyrme model to fix the moments of inertia and interaction potential. This was done for the standard Skyrme model as well as the loosely bound Skyrme model. The energy spectra of the free and full systems were calculated and compared to experimental data. Both spectra do include low-lying spin $1/2^-$ and $3/2^-$ states and a large gap. However the results contain approximate parity doubling. This is not seen experimentally. It is possible that one can remedy this problem by including the $D_2$-symmetric Skyrmion in the quantization scheme. Our work will serve as a foundation for this much more complicated problem, and many other two-cluster systems in the Skyrme model.

\subsection*{Acknowledgments}

We thank Chris King, Nick Manton and Jonathan Rawlinson for helpful discussions at the early stages of this project. CH would also like to thank Derek Harland for discussions on spin-orbit coupling within the Skyrme model. This work is supported by the National Natural Science Foundation of China (Grant No.~11675223). 

\appendix

\section{Basis wavefunctions} \label{app:A}

In this Appendix we tabulate bases of wavefunctions which transform as the irreps of the symmetry groups $\mathcal{G}_{JI}$ and $\mathcal{G}_{ST}$.  Tables \ref{wvreps1} and \ref{wvreps2} list sets of wavefunctions in the $J/I$ space, which transform as explained in Section \ref{sec:3}, for a variety of different $J/I$ values. For aesthetic reasons, it is convenient \textit{not} to normalize the wavefunctions. However, the wavefunctions in the text (used to construct the energy eigenstates) are assumed to be normalized. Since the symmetry groups are homomorphic, a table for the $\Phi_i$ wavefunctions on the $S/T$ space would be almost identical. In our conventions, the coefficients of the $S/T$  states are complex conjugates to those of the $J/I$ states.

\bgroup
\def \arraystretch{1.2}
\begin{table}[!ht]
	\begin{center}
		\begin{tabular}{| c | c | c | c | } \hline
			$(J, I)$ & Irrep. & $i$ & $\mathcal{N}\Ket{\Psi_i} = \sum a_{j_3 i_3}\Ket{J j_3}\Ket{I i_3}$  \\ \hline
			$\left(\frac{1}{2} \frac{1}{2}\right)$ & $\boldsymbol{4}$ & 1 & $\Ket{\frac{1}{2} \frac{1}{2}}\Ket{\frac{1}{2} \, {-\frac{1}{2}}}$ \\
			& & 2 & $-\Ket{\frac{1}{2} \frac{1}{2}}\Ket{\frac{1}{2} \frac{1}{2}}$ \\
			& & 3 & $-i\Ket{\frac{1}{2} \, {-\frac{1}{2}}}\Ket{\frac{1}{2} \, {-\frac{1}{2}}}$ \\
			& & 4 & $i \Ket{\frac{1}{2} \, {-\frac{1}{2}}}\Ket{\frac{1}{2} \frac{1}{2}}$ \\ \hline
			
			$\left(\frac{3}{2} \frac{1}{2}\right)$ & $\boldsymbol{4}$ & 1 & $\Ket{\frac{3}{2} \frac{1}{2}}\Ket{\frac{1}{2} \frac{1}{2}} + i \Ket{\frac{3}{2} \, {-\frac{3}{2}}}\Ket{\frac{1}{2} \frac{1}{2}}$ \\ 
			& & 2 & $-\Ket{\frac{3}{2} \frac{1}{2}}\Ket{\frac{1}{2} \, {-\frac{1}{2}}}+ i \Ket{\frac{3}{2} \, {-\frac{3}{2}}}\Ket{\frac{1}{2} \, {-\frac{1}{2}}}$ \\
			& & 3 & $-\Ket{\frac{3}{2} \frac{3}{2}}\Ket{\frac{1}{2} \frac{1}{2}} +i \Ket{\frac{3}{2} \, {-\frac{1}{2}}}\Ket{\frac{1}{2} \frac{1}{2}}$ \\
			& & 4 & $-\Ket{\frac{3}{2} \frac{3}{2}}\Ket{\frac{1}{2} \, {-\frac{1}{2}}} -i \Ket{\frac{3}{2} \, {-\frac{1}{2}}}\Ket{\frac{1}{2} \, {-\frac{1}{2}}}$ \\	
			& $\boldsymbol{2^+}$ & 1 & $\Ket{\frac{3}{2} \frac{3}{2}}\Ket{\frac{1}{2} \frac{1}{2}}+ \Ket{\frac{3}{2} \frac{3}{2}}\Ket{\frac{1}{2} \, {-\frac{1}{2}}}+i\Ket{\frac{3}{2} \, {-\frac{1}{2}}}\Ket{\frac{1}{2} \frac{1}{2}}- i \Ket{\frac{3}{2} \, {-\frac{1}{2}}}\Ket{\frac{1}{2} \, {-\frac{1}{2}}}$ \\
			&  & 2 & $\Ket{\frac{3}{2} \frac{1}{2}}\Ket{\frac{1}{2} \frac{1}{2}}- i\Ket{\frac{3}{2} \frac{1}{2}}\Ket{\frac{1}{2} \, {-\frac{1}{2}}}-i \Ket{\frac{3}{2} \, {-\frac{3}{2}}}\Ket{\frac{1}{2} \frac{1}{2}}-i\Ket{\frac{3}{2} \, {-\frac{3}{2}}}\Ket{\frac{1}{2} \, {-\frac{1}{2}}}$ \\
			& $\boldsymbol{2^-}$ & 1 & $\Ket{\frac{3}{2} \frac{3}{2}}\Ket{\frac{1}{2} \frac{1}{2}}- \Ket{\frac{3}{2} \frac{3}{2}}\Ket{\frac{1}{2} \, {-\frac{1}{2}}}+i\Ket{\frac{3}{2} \, {-\frac{1}{2}}}\Ket{\frac{1}{2} \frac{1}{2}}+ i \Ket{\frac{3}{2} \, {-\frac{1}{2}}}\Ket{\frac{1}{2} \, {-\frac{1}{2}}}$ \\
			&  & 2 & $\Ket{\frac{3}{2} \frac{1}{2}}\Ket{\frac{1}{2} \frac{1}{2}}+ i\Ket{\frac{3}{2} \frac{1}{2}}\Ket{\frac{1}{2} \, {-\frac{1}{2}}}-i\Ket{\frac{3}{2} \, {-\frac{3}{2}}}\Ket{\frac{1}{2} \frac{1}{2}}+i  \Ket{\frac{3}{2} \, {-\frac{3}{2}}}\Ket{\frac{1}{2} \, {-\frac{1}{2}}}$ \\ \hline
			
			$(\tfrac{5}{2} \tfrac{1}{2})$ & $\boldsymbol{4_1}$ & 1 & $\Ket{\frac{5}{2} \frac{5}{2}}\Ket{\frac{1}{2} \frac{1}{2}}$ \\
			& & 2 & $-\Ket{\frac{5}{2} \frac{1}{2}}\Ket{\frac{1}{2} \, {-\frac{1}{2}}}$ \\
			& & 3 & $-i\Ket{\frac{5}{2} \, {-\frac{1}{2}}}\Ket{\frac{1}{2} \frac{1}{2}}$  \\
			& & 4 & $i\Ket{\frac{5}{2} \, {-\frac{5}{2}}}\Ket{\frac{1}{2} \, {-\frac{1}{2}}}$ \\
			& $\boldsymbol{4_2}$ & 1 & $\Ket{\frac{5}{2} \frac{3}{2}}\Ket{\frac{1}{2} \, {-\frac{1}{2}}}$ \\
			& & 2 & $-\Ket{\frac{5}{2} \frac{3}{2}}\Ket{\frac{1}{2} \frac{1}{2}}$ \\
			& & 3 & $-i\Ket{\frac{5}{2} \, {-\frac{3}{2}}}\Ket{\frac{1}{2} \, {-\frac{1}{2}}}$  \\
			& & 4 & $i\Ket{\frac{5}{2} \, {-\frac{3}{2}}}\Ket{\frac{1}{2} \frac{1}{2}}$ \\
			
			& $\boldsymbol{2^+}$ & 1 & $\Ket{\frac{5}{2} \, {-\frac{1}{2}}}\Ket{\frac{1}{2} \, {-\frac{1}{2}}}-\Ket{\frac{5}{2} \, {-\frac{5}{2}}}\Ket{\frac{1}{2} \frac{1}{2}}$ \\
			& & 2 & $i\Ket{\frac{5}{2} \frac{5}{2}}\Ket{\frac{1}{2} \, {-\frac{1}{2}}}-i\Ket{\frac{5}{2} \frac{1}{2}}\Ket{\frac{1}{2} \frac{1}{2}}$ \\ 
			
			& $\boldsymbol{2^-}$ & 1 & $\Ket{\frac{5}{2} \, {-\frac{1}{2}}}\Ket{\frac{1}{2} \, {-\frac{1}{2}}}+\Ket{\frac{5}{2} \, {-\frac{5}{2}}}\Ket{\frac{1}{2} \frac{1}{2}}$ \\
			& & 2 & $i\Ket{\frac{5}{2} \frac{5}{2}}\Ket{\frac{1}{2} \, {-\frac{1}{2}}}+i\Ket{\frac{5}{2} \frac{1}{2}}\Ket{\frac{1}{2} \frac{1}{2}}$ \\ \hline
			
			$\left(\frac{1}{2} \frac{3}{2}\right)$ & $\boldsymbol{4}$ & 1 &  $\Ket{\frac{1}{2} \frac{1}{2}}\Ket{\frac{3}{2} \, {-\frac{1}{2}}}$ \\
			& & 2 & $\Ket{\frac{1}{2} \frac{1}{2}}\Ket{\frac{3}{2} \frac{1}{2}}$  \\
			& & 3 & $-i\Ket{\frac{1}{2} - \frac{1}{2}}\Ket{\frac{3}{2} \, {-\frac{1}{2}}}$ \\
			& & 4 & $-i\Ket{\frac{1}{2} - \frac{1}{2}}\Ket{\frac{3}{2} \frac{1}{2}}$ \\
			& $\boldsymbol{2^+}$ & 1 &$\Ket{\frac{1}{2} \, {-\frac{1}{2}}}\Ket{\frac{3}{2} \frac{3}{2}} + \Ket{\frac{1}{2} \, {-\frac{1}{2}}}\Ket{\frac{3}{2} \, {-\frac{3}{2}}}$  \\
			& & 2 & $i\Ket{\frac{1}{2} \frac{1}{2}}\Ket{\frac{3}{2} \frac{3}{2}} + i\Ket{\frac{1}{2} \frac{1}{2}}\Ket{\frac{3}{2} \, {-\frac{3}{2}}}$  \\
			 & $\boldsymbol{2^-}$ & 1 &$\Ket{\frac{1}{2} \, {-\frac{1}{2}}}\Ket{\frac{3}{2} \frac{3}{2}} - \Ket{\frac{1}{2} \, {-\frac{1}{2}}}\Ket{\frac{3}{2} \, {-\frac{3}{2}}}$  \\
			& & 2 & $i\Ket{\frac{1}{2} \frac{1}{2}}\Ket{\frac{3}{2} \frac{3}{2}} - i\Ket{\frac{1}{2} \frac{1}{2}}\Ket{\frac{3}{2} \, {-\frac{3}{2}}}$  \\
			
			\hline
		\end{tabular}
		\vskip 7pt
		\caption{Sets of spin states which transform under the irreps of the symmetry group $\mathcal{G}_{JI}$. These are chosen to transform as the realization of the irreps detailed in Section \ref{sec:4}. The factor of $\mathcal{N}$ represents the fact that the wavefunctions in the table are not normalized, though the ones used in the text are assumed to be.}
		\label{wvreps1}
	\end{center}
\end{table}
\egroup

\bgroup
\def \arraystretch{1.2}
\begin{table}[!ht]
	\begin{center}
		\begin{tabular}{ | c | c | c | } \hline
			 Irrep. & $i$ & $\mathcal{N}\Ket{\Psi_i} = \sum a_{j_3 i_3}\Ket{J j_3}\Ket{I i_3}$  \\ \hline
			 $\boldsymbol{4_1} $ & 1 & $\Ket{\frac{3}{2} \frac{1}{2}}\Ket{\frac{3}{2} \frac{3}{2}}-i\Ket{\frac{3}{2} \, {-\frac{3}{2}}}\Ket{\frac{3}{2} \frac{3}{2}}$ \\
			  & 2 & $\Ket{\frac{3}{2} \frac{1}{2}}\Ket{\frac{3}{2} \, {-\frac{3}{2}}}+i\Ket{\frac{3}{2} \, {-\frac{3}{2}}}\Ket{\frac{3}{2} \, {-\frac{3}{2}}}$ \\
			 & 3 & $\Ket{\frac{3}{2} \frac{3}{2}}\Ket{\frac{3}{2} \frac{3}{2}}+i\Ket{\frac{3}{2} \, {-\frac{1}{2}}}\Ket{\frac{3}{2} \frac{3}{2}}$ \\
			 & 4 & $-\Ket{\frac{3}{2} \frac{3}{2}}\Ket{\frac{3}{2} \, {-\frac{3}{2}}}+i\Ket{\frac{3}{2} \, {-\frac{1}{2}}}\Ket{\frac{3}{2} \, {-\frac{3}{2}}}$ \\
		 $\boldsymbol{4_2}$ & 1 & $\Ket{\frac{3}{2} \frac{1}{2}}\Ket{\frac{3}{2} \frac{1}{2}}+i\Ket{\frac{3}{2} \, {-\frac{3}{2}}}\Ket{\frac{3}{2} \frac{1}{2}}$ \\
			 & 2 & $\Ket{\frac{3}{2} \frac{1}{2}}\Ket{\frac{3}{2} \, {-\frac{1}{2}}}-i\Ket{\frac{3}{2} \, {-\frac{3}{2}}}\Ket{\frac{3}{2} \, {-\frac{1}{2}}}$ \\
			 & 3 & $-\Ket{\frac{3}{2} \frac{3}{2}}\Ket{\frac{3}{2} \frac{1}{2}}+i\Ket{\frac{3}{2} \frac{1}{2}}\Ket{\frac{3}{2} \frac{1}{2}}$ \\
			 & 4 & $\Ket{\frac{3}{2} \frac{3}{2}}\Ket{\frac{3}{2} \, {-\frac{1}{2}}}+i\Ket{\frac{3}{2} \, {-\frac{1}{2}}}\Ket{\frac{3}{2} \, {-\frac{1}{2}}}$ \\
			 $\boldsymbol{4_3}$ & 1 & $\Ket{\frac{3}{2} \frac{1}{2}}\Ket{\frac{3}{2} \, {-\frac{3}{2}}}-i\Ket{\frac{3}{2} \, {-\frac{3}{2}}}\Ket{\frac{3}{2} \, {-\frac{3}{2}}}$ \\
	 & 2 & $\Ket{\frac{3}{2} \frac{1}{2}}\Ket{\frac{3}{2} \frac{3}{2}}+i\Ket{\frac{3}{2} \, {-\frac{3}{2}}}\Ket{\frac{3}{2} \frac{3}{2}}$ \\
			 & 3 & $\Ket{\frac{3}{2} \frac{3}{2}}\Ket{\frac{3}{2} \, {-\frac{3}{2}}}+i\Ket{\frac{3}{2} \, {-\frac{1}{2}}}\Ket{\frac{3}{2} \, {-\frac{3}{2}}}$ \\
			 & 4 & $-\Ket{\frac{3}{2} \frac{3}{2}}\Ket{\frac{3}{2} \frac{3}{2}}+i\Ket{\frac{3}{2} \, {-\frac{1}{2}}}\Ket{\frac{3}{2} \frac{3}{2}}$ \\
			 $\boldsymbol{2^+}$ & 1 & $\Ket{\frac{3}{2} \frac{3}{2}}\Ket{\frac{3}{2} \frac{1}{2}}+\Ket{\frac{3}{2} \frac{3}{2}}\Ket{\frac{3}{2} \, {-\frac{1}{2}}}+i\Ket{\frac{3}{2} \, {-\frac{1}{2}}}\Ket{\frac{3}{2} \frac{1}{2}}- i\Ket{\frac{3}{2} \, {-\frac{1}{2}}}\Ket{\frac{3}{2} \, {-\frac{1}{2}}}$ \\
			 & 2 & $\Ket{\frac{3}{2} \frac{1}{2}}\Ket{\frac{3}{2} \frac{1}{2}}-\Ket{\frac{3}{2} \frac{1}{2}}\Ket{\frac{3}{2} \, {-\frac{1}{2}}}-i\Ket{\frac{3}{2} \, {-\frac{3}{2}}}\Ket{\frac{3}{2} \frac{1}{2}}- i\Ket{\frac{3}{2} \, {-\frac{3}{2}}}\Ket{\frac{3}{2} \, {-\frac{1}{2}}}$ \\
			 $\boldsymbol{2^-}$ & 1 & $\Ket{\frac{3}{2} \frac{3}{2}}\Ket{\frac{3}{2} \frac{1}{2}}-\Ket{\frac{3}{2} \frac{3}{2}}\Ket{\frac{3}{2} \, {-\frac{1}{2}}}+i\Ket{\frac{3}{2} \, {-\frac{1}{2}}}\Ket{\frac{3}{2} \frac{1}{2}}+i\Ket{\frac{3}{2} \, {-\frac{1}{2}}}\Ket{\frac{3}{2} \, {-\frac{1}{2}}}$ \\
			 & 2 & $\Ket{\frac{3}{2} \frac{1}{2}}\Ket{\frac{3}{2} \frac{1}{2}}+\Ket{\frac{3}{2} \frac{1}{2}}\Ket{\frac{3}{2} \, {-\frac{1}{2}}}-i\Ket{\frac{3}{2} \, {-\frac{3}{2}}}\Ket{\frac{3}{2} \frac{1}{2}}+ i\Ket{\frac{3}{2} \, {-\frac{3}{2}}}\Ket{\frac{3}{2} \, {-\frac{1}{2}}}$ \\ \hline
			 $\boldsymbol{4_1}$ & 1 & $\sqrt{3} \Ket{\frac{7}{2} \frac{5}{2}}\Ket{\frac{1}{2} \frac{1}{2}}+i \sqrt{5} \Ket{\frac{7}{2} \frac{1}{2}}\Ket{\frac{1}{2} \frac{1}{2}}+3\Ket{\frac{7}{2} \, {-\frac{3}{2}}}\Ket{\frac{1}{2} \frac{1}{2}}- i \sqrt{7}\Ket{\frac{7}{2} \, {-\frac{7}{2}}}\Ket{\frac{1}{2} \frac{1}{2}}$ \\
			 & 2 & $\sqrt{3} \Ket{\frac{7}{2} \frac{5}{2}}\Ket{\frac{1}{2} \, {-\frac{1}{2}}}-i \sqrt{5} \Ket{\frac{7}{2} \frac{1}{2}}\Ket{\frac{1}{2} \, {-\frac{1}{2}}}+3\Ket{\frac{7}{2} \, {-\frac{3}{2}}}\Ket{\frac{1}{2} \, {-\frac{1}{2}}}+ i \sqrt{7}\Ket{\frac{7}{2} \, {-\frac{7}{2}}}\Ket{\frac{1}{2} \, {-\frac{1}{2}}}$ \\
			 & 3 & $\sqrt{7} \Ket{\frac{7}{2} \frac{7}{2}}\Ket{\frac{1}{2} \frac{1}{2}}+3i  \Ket{\frac{7}{2} \frac{3}{2}}\Ket{\frac{1}{2} \frac{1}{2}}-\sqrt{5}\Ket{\frac{7}{2} \, {-\frac{1}{2}}}\Ket{\frac{1}{2} \frac{1}{2}}+ i \sqrt{3}\Ket{\frac{7}{2} \, {-\frac{5}{2}}}\Ket{\frac{1}{2} \frac{1}{2}}$ \\
			 & 4 & $-\sqrt{7} \Ket{\frac{7}{2} \frac{7}{2}}\Ket{\frac{1}{2} \, {-\frac{1}{2}}}+3i  \Ket{\frac{7}{2} \frac{3}{2}}\Ket{\frac{1}{2} \, {-\frac{1}{2}}}+\sqrt{5}\Ket{\frac{7}{2} \, {-\frac{1}{2}}}\Ket{\frac{1}{2} \, {-\frac{1}{2}}}+ i \sqrt{3}\Ket{\frac{7}{2} \, {-\frac{5}{2}}}\Ket{\frac{1}{2} \, {-\frac{1}{2}}}$ \\
			 $\boldsymbol{4_2}$ & 1 & $\sqrt{3} \Ket{\frac{7}{2} \frac{5}{2}}\Ket{\frac{1}{2} \, {-\frac{1}{2}}}-  \Ket{\frac{7}{2} \, {-\frac{3}{2}}}\Ket{\frac{1}{2} \, {-\frac{1}{2}}}$ \\
			  & 2 & $\sqrt{3} \Ket{\frac{7}{2} \frac{5}{2}}\Ket{\frac{1}{2} \frac{1}{2}}- \Ket{\frac{7}{2} \, {-\frac{3}{2}}}\Ket{\frac{1}{2} \frac{1}{2}}$ \\
			  & 3 & $-i  \Ket{\frac{7}{2} \frac{3}{2}}\Ket{\frac{1}{2} \, {-\frac{1}{2}}}+ i \sqrt{3} \Ket{\frac{7}{2} \, {-\frac{5}{2}}}\Ket{\frac{1}{2} \, {-\frac{1}{2}}}$ \\
			  & 4 & $-i  \Ket{\frac{7}{2} \frac{3}{2}}\Ket{\frac{1}{2} \frac{1}{2}}+ i \sqrt{3} \Ket{\frac{7}{2} \, {-\frac{5}{2}}}\Ket{\frac{1}{2} \frac{1}{2}}$ \\
			 $\boldsymbol{4_3}$ & 1 & $\sqrt{7} \Ket{\frac{7}{2} \frac{1}{2}}\Ket{\frac{1}{2} \, {-\frac{1}{2}}}-  \sqrt{5}\Ket{\frac{7}{2} \, {-\frac{7}{2}}}\Ket{\frac{1}{2} \, {-\frac{1}{2}}}$ \\
			 & 2 & $-\sqrt{7} \Ket{\frac{7}{2} \frac{1}{2}}\Ket{\frac{1}{2} \frac{1}{2}}-  \sqrt{5}\Ket{\frac{7}{2} \, {-\frac{7}{2}}}\Ket{\frac{1}{2} \frac{1}{2}}$ \\
			 & 3 & $i \sqrt{5} \Ket{\frac{7}{2} \frac{7}{2}}\Ket{\frac{1}{2} \, {-\frac{1}{2}}}+ i \sqrt{7}\Ket{\frac{7}{2} \, {-\frac{1}{2}}}\Ket{\frac{1}{2} \, {-\frac{1}{2}}}$ \\
			 & 4 & $-i \sqrt{5} \Ket{\frac{7}{2} \frac{7}{2}}\Ket{\frac{1}{2} \frac{1}{2}}- i \sqrt{7}\Ket{\frac{7}{2} \, {-\frac{1}{2}}}\Ket{\frac{1}{2} \frac{1}{2}}$ \\
			 $\boldsymbol{2^+}$ & 1 & $\sqrt{7} \Ket{\frac{7}{2} \frac{7}{2}}\Ket{\frac{1}{2} \frac{1}{2}}-\sqrt{7} \Ket{\frac{7}{2} \frac{7}{2}}\Ket{\frac{1}{2} \, {-\frac{1}{2}}} -3i\Ket{\frac{7}{2} \frac{3}{2}}\Ket{\frac{1}{2} \frac{1}{2}}-3i\Ket{\frac{7}{2} \frac{3}{2}}\Ket{\frac{1}{2} \, {-\frac{1}{2}}} - $ \\
			 & & $\sqrt{5} \Ket{\frac{7}{2} \, {-\frac{1}{2}}}\Ket{\frac{1}{2} \frac{1}{2}}+\sqrt{5}\Ket{\frac{7}{2} \, {-\frac{1}{2}}}\Ket{\frac{1}{2} \, {-\frac{1}{2}}} - i \sqrt{3}\Ket{\frac{7}{2} \, {-\frac{5}{2}}}\Ket{\frac{1}{2} \frac{1}{2}}- i \sqrt{3}\Ket{\frac{7}{2} \, {-\frac{5}{2}}}\Ket{\frac{1}{2} \, {-\frac{1}{2}}} $ \\
			& 2 & $-\sqrt{3} \Ket{\frac{7}{2} \frac{5}{2}}\Ket{\frac{1}{2} \frac{1}{2}}-\sqrt{3} \Ket{\frac{7}{2} \frac{5}{2}}\Ket{\frac{1}{2} \, {-\frac{1}{2}}}+i\sqrt{5}  \Ket{\frac{7}{2} \frac{1}{2}}\Ket{\frac{1}{2} \frac{1}{2}}-i\sqrt{5}  \Ket{\frac{7}{2} \frac{1}{2}}\Ket{\frac{1}{2} \, {-\frac{1}{2}}} -$ \\
			& & $3 \Ket{\frac{7}{2} \, {-\frac{3}{2}}}\Ket{\frac{1}{2} \frac{1}{2}} -3\Ket{\frac{7}{2} \, {-\frac{3}{2}}}\Ket{\frac{1}{2} \, {-\frac{1}{2}}}-i \sqrt{7}\Ket{\frac{7}{2} \, {-\frac{7}{2}}}\Ket{\frac{1}{2} \frac{1}{2}} +i \sqrt{7}\Ket{\frac{7}{2} \, {-\frac{7}{2}}}\Ket{\frac{1}{2} \, {-\frac{1}{2}}} $ \\
			$\boldsymbol{2^-}$ & 1 & $\sqrt{7} \Ket{\frac{7}{2} \frac{7}{2}}\Ket{\frac{1}{2} \frac{1}{2}} + \sqrt{7} \Ket{\frac{7}{2} \frac{7}{2}}\Ket{\frac{1}{2} \, {-\frac{1}{2}}} - 3i \Ket{\frac{7}{2} \frac{3}{2}}\Ket{\frac{1}{2} \frac{1}{2}}+ 3i \Ket{\frac{7}{2} \frac{3}{2}}\Ket{\frac{1}{2} \, {-\frac{1}{2}}}-$ \\ & & $ \sqrt{5} \Ket{\frac{7}{2} \, {-\frac{1}{2}}}\Ket{\frac{1}{2} \frac{1}{2}}-\sqrt{5} \Ket{\frac{7}{2} \, {-\frac{1}{2}}}\Ket{\frac{1}{2} \, {-\frac{1}{2}}} - i \sqrt{3}  \Ket{\frac{7}{2} \, {-\frac{5}{2}}}\Ket{\frac{1}{2} \frac{1}{2}}+i \sqrt{3}  \Ket{\frac{7}{2} \, {-\frac{5}{2}}}\Ket{\frac{1}{2} \, {-\frac{1}{2}}}$ \\
			 & 2 & $-\sqrt{3}\Ket{\frac{7}{2} \frac{5}{2}}\Ket{\frac{1}{2} \frac{1}{2}}+\sqrt{3}\Ket{\frac{7}{2} \frac{5}{2}}\Ket{\frac{1}{2} \, {-\frac{1}{2}}} +i\sqrt{5}\Ket{\frac{7}{2} \frac{1}{2}}\Ket{\frac{1}{2} \frac{1}{2}} +i\sqrt{5}\Ket{\frac{7}{2} \frac{1}{2}}\Ket{\frac{1}{2} \, {-\frac{1}{2}}}-$ \\ & & $3 \Ket{\frac{7}{2} \, {-\frac{3}{2}}}\Ket{\frac{1}{2} \frac{1}{2}}+3\Ket{\frac{7}{2} \, {-\frac{3}{2}}}\Ket{\frac{1}{2} \, {-\frac{1}{2}}}-i \sqrt{7}\Ket{\frac{7}{2} \, {-\frac{7}{2}}}\Ket{\frac{1}{2} \frac{1}{2}}-i \sqrt{7}\Ket{\frac{7}{2} \, {-\frac{7}{2}}}\Ket{\frac{1}{2}  \, {-\frac{1}{2}}}$ \\
			\hline
		\end{tabular}
		\vskip 7pt
		\caption{More spin state bases, for $(J,I) = (3/2,3/2)$ and $(J,I) = (7/2,1/2)$, which transform under the irreps of $\mathcal{G}_{JI}$.}
		\label{wvreps2}
	\end{center}
\end{table}
\egroup

A permissible wavefunction takes the form
\begin{equation}
\Ket{\Psi} = \sum_i \Phi_i \Ket{\Psi_i} \, ,
\end{equation}
where the two sets of wavefunctions fall into the same irrep. For example, the $\Ket{\frac{3}{2} \frac{1}{2} \frac{1}{2} \frac{3}{2},\boldsymbol{2^+}}$ wavefunction is
\begin{align*}
\Ket{\tfrac{3}{2} \tfrac{1}{2} \tfrac{1}{2} \tfrac{3}{2} ; \boldsymbol{2^+}}= &\left( \Ket{\tfrac{3}{2}\, \tfrac{3}{2}}\Ket{\tfrac{1}{2}\, \tfrac{1}{2}}+ \Ket{\tfrac{3}{2}\, \tfrac{3}{2}}\Ket{\tfrac{1}{2}\, {-\tfrac{1}{2}}}+i\Ket{\tfrac{3}{2}\, {-\tfrac{1}{2}}}\Ket{\tfrac{1}{2}\, \tfrac{1}{2}}- i \Ket{\tfrac{3}{2}\, {-\tfrac{1}{2}}}\Ket{\tfrac{1}{2}\, {-\tfrac{1}{2}}} \right)\\ & \times \left(\Ket{\tfrac{1}{2}\, {-\tfrac{1}{2}}}\Ket{\tfrac{3}{2}\, \tfrac{3}{2}} + \Ket{\tfrac{1}{2}\, {-\tfrac{1}{2}}}\Ket{\tfrac{3}{2}\, {-\tfrac{3}{2}}}\right) \\
 &+ \left(\Ket{\tfrac{3}{2}\, \tfrac{1}{2}}\Ket{\tfrac{1}{2}\, \tfrac{1}{2}}- i\Ket{\tfrac{3}{2}\, \tfrac{1}{2}}\Ket{\tfrac{1}{2}\, {-\tfrac{1}{2}}}-i \Ket{\tfrac{3}{2}\, {-\tfrac{3}{2}}}\Ket{\tfrac{1}{2}\, \tfrac{1}{2}}-i\Ket{\tfrac{3}{2}\, {-\tfrac{3}{2}}}\Ket{\tfrac{1}{2}\, {-\tfrac{1}{2}}}\right)\\ &\times\left(- i\Ket{\tfrac{1}{2}\, \tfrac{1}{2}}\Ket{\tfrac{3}{2}\, \tfrac{3}{2}} - i\Ket{\tfrac{1}{2}\, \tfrac{1}{2}}\Ket{\tfrac{3}{2}\, {-\tfrac{3}{2}}} \right) \,  .
\end{align*}
We have suppressed the space-fixed projection for aesthetic reasons. We can then use these wavefunctions to generate the energy eigenfunctions.

\section{Energy eigenstates} \label{app:B}

Low energy eigenstates were mentioned in the main text. We tabulate several higher-energy states, including all of those used in Fig. 7, in tables \ref{energies}. We list them in order of increasing energy. In this table we assume that the constituent states have been normalized. 

In the search for low energy states we considered a wide variety of different $(J,I,S,T)$ combinations. For $(J,I)=(1/2,1/2)$ and $(J,I)=(3/2,1/2)$ we considered
\begin{equation}
(S,T) = \left(\left(\tfrac{1}{2},\tfrac{1}{2}\right),\left(\tfrac{3}{2},\tfrac{1}{2}\right),\left(\tfrac{5}{2},\tfrac{1}{2}\right),\left(\tfrac{7}{2},\tfrac{1}{2}\right),\left(\tfrac{1}{2},\tfrac{3}{2}\right),\left(\tfrac{3}{2},\tfrac{3}{2}\right),\left(\tfrac{1}{2},\tfrac{5}{2}\right)\right) \, .
\end{equation}
For the higher spins and isospins
\begin{equation}
(J,I) = \left(\left(\tfrac{5}{2},\tfrac{1}{2}\right),\left(\tfrac{7}{2},\tfrac{1}{2}\right),\left(\tfrac{1}{2},\tfrac{3}{2}\right),\left(\tfrac{3}{2},\tfrac{3}{2}\right)\right) \, ,
\end{equation}
we only considered $(S,T)=(1/2,1/2)$ and $(S,T)=(3/2,1/2)$. In total we calculated 144 energy eigenstates of the free system.

\bgroup
\def \arraystretch{1.4}
\begin{table}[!ht]
	\begin{center}
		\begin{tabular}{| c | c  |} \hline
			Energy eigenstate & Energy  \\ \hline
			
			$\Ket{\tfrac{1}{2}  \tfrac{1}{2} \tfrac{1}{2} \tfrac{1}{2};\boldsymbol{4}}$ & $\tfrac{3}{8v_1} +\tfrac{3}{4\mu}$  \\ \hline
			
			$\sqrt{2}\Ket{\tfrac{3}{2}  \tfrac{1}{2} \tfrac{3}{2} \tfrac{1}{2};\boldsymbol{4}} + \Ket{\tfrac{3}{2}  \tfrac{1}{2} \tfrac{3}{2} \tfrac{1}{2};\boldsymbol{2^+}}+\Ket{\tfrac{3}{2}  \tfrac{1}{2} \tfrac{3}{2} \tfrac{1}{2};\boldsymbol{2^-}}$ & $\tfrac{3}{8v_1}+\tfrac{9}{4\mu}$  \\ \hline
			
			$\Ket{\tfrac{3}{2}  \tfrac{1}{2} \tfrac{1}{2} \tfrac{1}{2};\boldsymbol{4}}$ & $\tfrac{3}{8v_1} +\tfrac{1}{2u_{11}}+\tfrac{1}{2u_{33}}+\tfrac{3}{v_4} +\tfrac{3}{4\mu}$ \\ 
			
			$\Ket{\tfrac{5}{2}  \tfrac{1}{2} \tfrac{1}{2} \tfrac{1}{2};\boldsymbol{4_1}}$ & $\tfrac{3}{8v_1} +\tfrac{1}{2u_{11}}+\tfrac{1}{2u_{33}}+\tfrac{3}{v_4} +\tfrac{3}{4\mu}$  \\
			$\Ket{\tfrac{3}{2}  \tfrac{3}{2} \tfrac{1}{2} \tfrac{1}{2};\boldsymbol{4_2}} - \sqrt{3}\Ket{\tfrac{3}{2}  \tfrac{3}{2} \tfrac{1}{2} \tfrac{1}{2};\boldsymbol{4_3}}$ & $\tfrac{3}{8v_1}+\tfrac{1}{2u_{11}}+\tfrac{1}{2u_{33}}+\tfrac{3}{v_4}+\tfrac{3}{4\mu}$ \\ \hline
			 
			$\Ket{\tfrac{1}{2}  \tfrac{1}{2} \tfrac{3}{2} \tfrac{1}{2};\boldsymbol{4}}$ & $\tfrac{3}{8v_1} +\tfrac{1}{2u_{11}}+\frac{1}{2u_{33}}+\tfrac{3}{v_4} +\tfrac{9}{4\mu}$ \\ 	
			$\Ket{\tfrac{3}{2}  \tfrac{1}{2} \tfrac{3}{2} \tfrac{1}{2};\boldsymbol{2^+}}-\Ket{\tfrac{3}{2}  \tfrac{1}{2} \tfrac{3}{2} \tfrac{1}{2};\boldsymbol{2^-}} $ & $\tfrac{3}{8v_1}+\tfrac{1}{2u_{11}}+\tfrac{1}{2u_{33}}+\tfrac{3}{v_4}+\tfrac{9}{4\mu}$ \\
			
			$ i \sqrt{5} \Ket{\tfrac{5}{2}  \tfrac{1}{2} \tfrac{3}{2} \tfrac{1}{2};\boldsymbol{4_2}}-\Ket{\tfrac{5}{2}  \tfrac{1}{2} \tfrac{3}{2} \tfrac{1}{2};\boldsymbol{2^+}}+\Ket{\tfrac{5}{2}  \tfrac{1}{2} \tfrac{3}{2} \tfrac{1}{2};\boldsymbol{2^-}}$ & $\tfrac{3}{8v_1}+\tfrac{1}{2u_{11}}+\tfrac{1}{2u_{33}}+\tfrac{3}{v_4}+\tfrac{9}{4\mu}$\\
			$\Big( \sqrt{5} \Ket{\tfrac{7}{2}  \tfrac{1}{2} \tfrac{3}{2} \tfrac{1}{2};\boldsymbol{4_2}} + i\sqrt{14}  \Ket{\tfrac{7}{2}  \tfrac{1}{2} \tfrac{3}{2} \tfrac{1}{2};\boldsymbol{4_3}}  - $  & \\
			$ \sqrt{10}\Ket{\tfrac{7}{2}  \tfrac{1}{2} \tfrac{3}{2} \tfrac{1}{2};\boldsymbol{2^+}} + \sqrt{10}\Ket{\tfrac{7}{2}  \tfrac{1}{2} \tfrac{3}{2} \tfrac{1}{2};\boldsymbol{2^-}}\Big)$ & $\tfrac{3}{8v_1}+\tfrac{1}{2u_{11}}+\tfrac{1}{2u_{33}}+\tfrac{3}{v_4}+\tfrac{9}{4\mu}$  \\
			$\sqrt{2}\Ket{\tfrac{1}{2}  \tfrac{3}{2} \tfrac{3}{2} \tfrac{1}{2};\boldsymbol{4}}+\sqrt{3}\Ket{\tfrac{1}{2}  \tfrac{3}{2} \tfrac{3}{2} \tfrac{1}{2};\boldsymbol{2^+}}+\sqrt{3}\Ket{\tfrac{1}{2}  \tfrac{3}{2} \tfrac{3}{2} \tfrac{1}{2};\boldsymbol{2^-}}$ & 
			$\tfrac{3}{8v_1}+\tfrac{1}{2u_{11}}+\tfrac{1}{2u_{33}}+\tfrac{3}{v_4}+\tfrac{9}{4\mu}$  \\ \hline
			$\Ket{\tfrac{1}{2}  \tfrac{1}{2} \tfrac{5}{2} \tfrac{1}{2};\boldsymbol{4_2}}$ & $\tfrac{3}{8v_1} +\tfrac{1}{2u_{11}}+\tfrac{1}{2u_{33}}+\tfrac{3}{v_4} +\tfrac{19}{4\mu}$   \\ 	
			$i\sqrt{5}\Ket{\tfrac{3}{2}  \tfrac{1}{2} \tfrac{5}{2} \tfrac{1}{2};\boldsymbol{4_2}}- \Ket{\tfrac{3}{2}  \tfrac{1}{2} \tfrac{5}{2} \tfrac{1}{2};\boldsymbol{2^+}}+\Ket{\tfrac{3}{2}  \tfrac{1}{2} \tfrac{5}{2} \tfrac{1}{2};\boldsymbol{2^-}}$ & $\tfrac{3}{8v_1}+\tfrac{1}{2u_{11}}+\tfrac{1}{2u_{33}}+\tfrac{3}{v_4}+\tfrac{19}{4\mu}$  \\ \hline
			$\Ket{\tfrac{7}{2}  \tfrac{1}{2} \tfrac{1}{2} \tfrac{1}{2};\boldsymbol{4_3}}$ & $\tfrac{3}{8v_1}+\tfrac{10}{v_4}+\tfrac{3}{4\mu} $ \\ \hline
			$\Ket{\tfrac{3}{2}  \tfrac{3}{2} \tfrac{3}{2} \tfrac{1}{2};\boldsymbol{4_1}}$ & $\tfrac{3}{4v_1} + \tfrac{1}{u_{11}}+\tfrac{1}{2u_{33}}+\tfrac{3}{v_4} + \tfrac{9}{4\mu}$  \\ \hline
			$\Ket{\tfrac{5}{2}  \tfrac{1}{2} \tfrac{1}{2} \tfrac{1}{2};\boldsymbol{4_1}}$ & $\tfrac{3}{4v_1} + \tfrac{1}{u_{11}}+\tfrac{6}{v_4} + \tfrac{3}{4\mu}$  \\ 
			$\Ket{\tfrac{7}{2}  \tfrac{1}{2} \tfrac{1}{2} \tfrac{1}{2};\boldsymbol{4_2}}$ & $\tfrac{3}{4v_1} + \tfrac{1}{u_{11}}+\tfrac{6}{v_4} + \tfrac{3}{4\mu}$  \\ \hline
			$-i \sqrt{2} \Ket{\tfrac{5}{2}  \tfrac{1}{2} \tfrac{3}{2} \tfrac{1}{2};\boldsymbol{4_1}}+\Ket{\tfrac{5}{2}  \tfrac{1}{2} \tfrac{3}{2} \tfrac{1}{2};\boldsymbol{2^+}}+\Ket{\tfrac{5}{2}  \tfrac{1}{2} \tfrac{3}{2} \tfrac{1}{2};\boldsymbol{2^-}}$ & $\tfrac{3}{8v_1}+\tfrac{10}{v_4}+\tfrac{9}{4\mu}$\\ 
			 $ \sqrt{2} \Ket{\tfrac{7}{2}  \tfrac{1}{2} \tfrac{3}{2} \tfrac{1}{2};\boldsymbol{4_1}}+\Ket{\tfrac{7}{2}  \tfrac{1}{2} \tfrac{3}{2} \tfrac{1}{2};\boldsymbol{2^+}}+\Ket{\tfrac{7}{2}  \tfrac{1}{2} \tfrac{3}{2} \tfrac{1}{2};\boldsymbol{2^-}}$ & $\tfrac{3}{8v_1}+\tfrac{10}{v_4}+\tfrac{9}{4\mu} $\\
			 \hline
		\end{tabular}
		\vskip 7pt
		\caption{The 19 lowest energy eigenfunctions and their corresponding energies. We use the notation $\Ket{J \, I \, S \, T \, ; \boldsymbol{X_i}}$ as described in Section \ref{sec:4}. We omit $t_3'$ as this has no effect on the energy.}
		\label{energies}
	\end{center}
\end{table}
\egroup

\end{document}